\documentclass[aps,prc,reprint,amsmath]{revtex4-1}

\usepackage{bm,graphicx,dcolumn}

\newcommand\etal{\textit{~et~al.}}
\renewcommand\bar\overline

\begin{document}

\title{Role of pair-vibrational correlations in forming the odd-even
  mass difference}

\author{K. Neerg\aa rd}
\affiliation{Fjordtoften 17, 4700 N\ae stved, Denmark}

\author{I. Bentley}
\affiliation{Department of Chemistry and Physics,
  Saint Mary's College, Notre Dame, Indiana 46556, USA}

\begin{abstract}

  In the random-phase-approximation-amended (RPA-amended)
  Nilsson-Strutinskij method of calculating nuclear binding energies,
  the conventional shell correction terms derived from the
  independent-nucleon model and the Bardeen-Cooper-Schrieffer pairing
  theory are supplemented by a term which accounts for the
  pair-vibrational correlation energy. This term is derived by means
  of the RPA from a pairing Hamiltonian which includes a
  neutron-proton pairing interaction. The method was used previously
  in studies of the pattern of binding energies of nuclei with
  approximately equal numbers $N$ and $Z$ of neutrons and protons and
  even mass number $A = N + Z$. Here it is applied to odd-$A$ nuclei.
  Three sets of such nuclei are considered: (i) The sequence of nuclei
  with $Z = N - 1$ and $25 \le A \le 99$. (ii) The odd-$A$ isotopes of
  In, Sn, and Sb with $46 \le N \le 92$. (iii) The odd-$A$ isotopes of
  Sr, Y, Zr, Nb, and Mo with $60 \le N \le 64$. The RPA correction is
  found to contribute significantly to the calculated odd-even mass
  differences, particularly in the light nuclei. In the upper
  \textit{sd} shell this correction accounts for almost the entire
  odd-even mass difference for odd $Z$ and about half of it for odd
  $N$. The size and sign of the RPA contribution varies, which is
  explained qualitatively in terms of a closed expression for a smooth
  RPA counter term.

\end{abstract}

\maketitle

\section{\label{sec:intr}Introduction}

Nuclear binding energies are often calculated in mean-field
approximations. The Bardeen-Cooper-Schrieffer (BCS) theory of
superconductivity~\cite{ref:Bar57a,*ref:Bar57b}, which was applied
extensively to the description of pairing in nuclei since its adaption
to the nuclear system by Bohr, Mottelson, and Pines~\cite{ref:Boh58},
Bogolyubov~\cite{ref:Bog58}, and
Solov'yov~\cite{ref:Sol58a,*ref:Sol58b}, is such an approximation.
Residual interactions, which are neglected in a mean-field
approximation, induce \textit{correlations}, which increase the
binding energy. We call this extra binding energy \textit{correlation
  energy} (in Ref.~\cite{ref:Ban70} this term is used differently.)
The BCS theory, in particular, may be derived, for a given type of
fermion (electron, neutron, proton), from the Hamiltonian
\begin{equation}\label{eq:HBCS}
  H = \sum_k \epsilon_k a_k^\dagger a_k - G P^\dagger P , \quad
  P = \tfrac12 \sum_k a_{\bar k} a_k .
\end{equation}
Here $a_k$ annihilates a fermion in a member $| k \rangle$ of an
orthonormal set of single-fermion states which is preserved up to
phases under time reversal, denoted by the bar. The single-fermion
energies $\epsilon_k = \epsilon_{\bar k}$ and the coupling constant
$G$ are parameters. The second term in the expression~\eqref{eq:HBCS}
is known as the \textit{pairing interaction}. The exact minimum of the
Hamiltonian~\eqref{eq:HBCS} can be calculated with any wanted accuracy
for fairly large single-fermion spaces~\cite{ref:Ric63}.
\begin{figure}{\centering\includegraphics{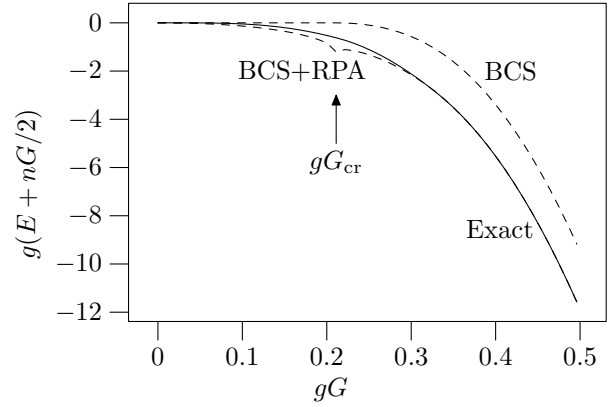}\par}
  \caption{\label{fig:bang}Adapted from Fig.~1 of
    Ref.~\cite{ref:Ban70}. The exact minimum $E$ of the 
    Hamiltonian~\eqref{eq:HBCS}, normalized to zero for $G = 0$, is
    shown as a function of $G$ in comparison with the approximations
    BCS and BCS+RPA. The single-fermion space accommodates 32
    equidistant doublet levels $\epsilon_k = \epsilon_{\bar k}$ spaced
    by $1/g$ and is inhabited by $n = 32$
    fermions. The expectation value $- G n/2$ of the pairing
    interaction in the $G = 0$ ground state is subtracted from the
    exact and RPA energies. The threshold $G_\text{cr}$ of BCS pairing
    is indicated. We turned the figure upside down to display energy
    rather than binding energy.}
\end{figure}
Figure~\ref{fig:bang} shows the result of such a calculation in
comparison with that obtained when the correlation energy is
calculated in the random phase approximation (RPA)~\cite{ref:Bom53}.
This approximation is seen to give a good agreement with the exact
value. Appreciable deviations only occur in a narrow interval of $G$
about the threshold $G_\text{cr}$ of BCS pairing. Because the RPA
equations derived from the Hamiltonian~\eqref{eq:HBCS} describe
oscillations of the pair field $P$ about the mean field equilibrium,
the correlations may thus be seen as mainly \textit{pair vibrational}.

Calculations of binding energies by the Strutinskij
method~\cite{ref:Str66} conventionally include a pairing term based on
the BCS theory. Figure~\ref{fig:bang} indicates a significance of the
correlation energy which suggests that it be taken into account. For
$G < G_\text{cr}$, in particular, the pairing interaction induces only
correlation energy. Moreover, isobaric invariance requires that the
sum of neutron and proton pairing interactions be generalized to
\begin{equation}\label{eq:ivecp}
   - G \vec P^\dagger \cdot \vec P
\end{equation}
with a pair field isovector
\begin{equation}\label{eq:vecP}
  \vec P = i \sqrt2 \sum_{kl}
    \langle \bar l | t_y \vec t \, | k \rangle a_l a_k . 
\end{equation}
Here $\vec t = (t_x , t_y , t_z)$ is the single-nucleon isospin, and
time reversal is assumed to commute with $t_x$ and $t_z$ and
anticommute with $t_y$. In Eq.~\eqref{eq:vecP} the set $k$ or $l$ of
quantum numbers includes an eigenvalue of $t_z$, and the span of the
orthonormal set of states $|k \rangle$ is isobarically invariant. The
interaction~\eqref{eq:ivecp} contains a neutron-proton term
$- G P_z^\dagger P_z$. In a doubly even nucleus the Hartree-Bogolyubov
quasinucleon vacuum derived from the resulting Hamiltonian has
$\langle P_z \rangle = 0$~\cite{ref:Nee09}, so the neutron-proton
interaction also induces only correlation energy.

In a collaboration with Frauendorf we developed an extension of the
conventional Nilsson-Strutinskij scheme which takes the
pair-vibrational correlations into account in the
RPA~\cite{ref:Ben14}. Minor modifications of the scheme of
calculations proposed in Ref.~\cite{ref:Ben14} were discussed by
Neerg\aa rd~\cite{ref:Nee16,ref:Nee17}. These articles deal with
nuclei with $N \approx Z$ and even $A$, where $N$ and $Z$ are the
numbers of neutrons and protons and $A = N + Z$. The extended
Nilsson-Strutinskij scheme was found to account, with suitably chosen
parameters, quite well for the pattern of even-$A$ binding energies
and certain excitation energies in doubly odd nuclei in this region.
We here apply it to odd-$A$ nuclei. We examine in particular the
influence of the inclusion of the RPA term on the calculated odd-even
mass differences. Three regions of the chart of nuclei are considered:
(i) The $N \approx Z$ region, previously studied with respect to the
even-$A$ nuclei. (ii) A neighborhood of the Sn isotopic chain. (iii) A
region of well-deformed, neutron rich nuclei around $^{102}$Zr.

The organization of the article is as follows. In Sec.~\ref{sec:mod}
we describe the scheme of calculations. This section serves to present
in one place all ingredients of the RPA-amended Nilsson-Strutinskij
method in the form it has taken after several modifications since the
publication of Ref.~\cite{ref:Ben14}. Then, in each of
Secs.~\ref{sec:N=Z}--\ref{sec:Zr}, we discuss the results for one of
the regions (i)--(iii). Finally, after exploring in Sec.~\ref{sec:int}
a technical matter of interpolation of the RPA energy across the
threshold of BCS pairing, we summarize our results in
Sec.~\ref{sec:sum}.

\section{\label{sec:mod}RPA-amended Nilsson-Strutinskij model}

The binding energy $- E(N,Z)$ is calculated by
\begin{multline}\label{eq:E}
  E(N,Z) = E_\text{LD} \\
    +  \sum_{\tau = n, p}
      ( \delta E_\text{i.n.,$\tau$} + \delta E_\text{BCS,$\tau$} )
    + \hskip-.5em \sum_{\tau = n, p, np} \hskip-.5em
      \delta E_\text{RPA,$\tau$} ,
\end{multline}
where `i.n.' stands for `independent nucleons'. Here $E_\text{LD}$ is
a liquid drop energy, and each term $\delta E_x$ has the form
\begin{equation}
   \delta E_x = E_x - \tilde E_x
\end{equation}
with a `smooth' counter term $\tilde E_x$. The `microscopic' energy
\begin{equation}\label{eq:mic}
  E_\text{mic}
  = \sum_{\tau = n, p} ( E_\text{i.n.,$\tau$} + E_\text{BCS,$\tau$} )
    + \hskip-.5em \sum_{\tau = n, p, np} \hskip-.5em E_\text{RPA,$\tau$}
\end{equation}
approximates the minimum of the Hamiltonian
\begin{equation}\label{eq:H}
  H = \sum_{\tau = n, p} \sum_{k = 1}^{2 \Omega_\tau}
      \epsilon_{k \tau} a_{k \tau}^\dagger a_{k \tau}
    - \hskip-.5em \sum_{\tau = n, p, np} \hskip-.5em
      G_\tau P_\tau^\dagger P_\tau ,
\end{equation}
where
\begin{equation}\label{eq:P}\begin{gathered}
  P_n = \tfrac12 \sum_{k = 1}^{2 \Omega_n} a_{\bar{k n}} a_{k n} , \\
  P_p = \tfrac12 \sum_{k = 1}^{2 \Omega_p} a_{\bar{k p}} a_{k p} , \\
  P_{np} = 2^{-\frac32} \sum_{k = 1}^{2 \Omega_{np}}
    ( a_{\bar{k p}} a_{k n} + a_{\bar{k n}} a_{k p} ) .
\end{gathered}\end{equation}
Here, unlike in Eq.~\eqref{eq:vecP}, the index $k$ numbers, for each
$\tau = n$ for neutrons and $\tau = p$ for protons, an orthonormal set
of eigenstates $| k \tau \rangle$ of a time-reversal invariant
single-nucleon Hamiltonian $h_\tau$ in an order of nondecreasing
eigenvalue $\epsilon_{k \tau}$. The numbering should be such that
$| k p \rangle = t_-| k n \rangle$ in the limit $h_p = h_n$. In this
limit then $P_n = - P_- / \sqrt2$, $P_p = P_+ / \sqrt2$, and
$P_{np} = P_z$ in terms of components of the isovector~\eqref{eq:vecP}
provided also all $\Omega_\tau$ are equal. Again the set of states
$| k \tau \rangle$ is supposed to be preserved under time reversal up
to phases. We also assume that each pair of an odd and the following
even $k$ refer to a pair of states connected by time reversal up to
phases. Both of these assumptions are satisfied automatically if the
eigenvalues are doubly degenerate, that is, except in spherical
nuclei. In the spherical case it is satisfied if degenerate orbits are
distinguished by a magnetic quantum number $m$ and pairs of an odd and
the following even $k$ refer to pairs of states with opposite $m$.

Unlike Ref.~\cite{ref:Ben14} strict isobaric invariance is not imposed
on the microscopic model. The single-nucleon Hamiltonians $h_n$ and
$h_p$ may be different, and different valence space dimension
$2 \Omega_\tau$ may be employed for different $\tau$. We use
throughout $\Omega_n = N$, $\Omega_p = Z$, and
$\Omega_{np} = \lceil A/2 \rceil$ so that the neutron and proton
valence spaces are always half filled and
$\Omega_{np} \approx (\Omega_n + \Omega_p)/2$. These modifications,
which where introduced partly in Refs.~\cite{ref:Nee16,ref:Nee17},
renders the model better suited for nuclei with a large
neutron or proton excess.

We also allow different coupling constants $G_\tau$ for different
$\tau$, writing
\begin{equation}\label{eq:G}
  G_\tau = \bar G A^\zeta (1 - \alpha M_T M_T') ,
\end{equation}
where $M_T = (N - Z)/2$ is the isomagnetic quantum number of the
nucleus and $M_T'$ that of the interacting pair, that is, $M_T' = 1$,
$-1$ and 0 for $\tau = n$, $p$ and $np$, respectively. The
parameters $\bar G$, $\zeta$, and $\alpha$ are set separately for each
region (i)--(iii). The limit where $h_p =h_n$, all $G_\tau$ are equal,
and all $\Omega_\tau$ are equal will be referred to as the
\textit{limit of isobaric invariance}.

For each nucleus we assume a deformation, which we take from a
conventional Nilsson-Strutinskij calculation~\cite{ref:Ben10}. It is
expressed by the Nilsson parameters $\epsilon_2$, $\gamma$, and
$\epsilon_4$~\cite{ref:Nil69,ref:Lar73}. The deformations are listed
in the appendix.

\subsection{\label{sec:LD}Liquid drop energy}

The liquid drop energy is written
\begin{multline}\label{eq:LD}
  E_\text{LD} =
     - \left(a_v - a_{vt} \frac{|M_T|(|M_T|+1)}{A^2} \right) A \\
     + \left(a_s - a_{st} \frac{|M_T|(|M_T|+1)}{A^2} \right) A^{2/3} B_s
     + a_c \frac{Z(Z-1)}{A^{1/3}} B_c ,
\end{multline}
where the coefficients $a_x$ are parameters. The deformation
dependent factors $B_s$ and $B_c$ are calculated from the Nilsson
parameters in two steps. First, following Seeger and
Howard~\cite{ref:See75}, we determine the coefficients $\alpha_{lm}$
in the equations in spherical coordinates $(r,\theta,\phi)$ of the
surfaces of constant second term in the expression~\eqref{eq:hNil}
below,
\begin{equation}\label{eq:Leg}
  r \propto 1 + \hskip-.7em \sum_{|m| \le l > 0} \hskip-.7em (-)^m
    \alpha_{lm} \sqrt{\frac{(l-|m|)!}{(l+|m|)!}}
    P_l^{|m|}(\cos \theta) \exp ( - i m \phi ) ,
\end{equation}
where $P_l^m(x)$ is the Legendre function of the first kind as defined
by Edmonds~\cite{ref:Edm57}. With
$\epsilon_{20} = \epsilon_2 \cos \gamma$ and
$\epsilon_{22} = (- \epsilon_2 \sin \gamma) / \sqrt2$, the nonzero
coefficients with $l \le 4$ are given to second order in $\epsilon_2$
and $\epsilon_4$ by
\begin{equation}\label{eq:ep>al}\begin{gathered}
  \alpha_{20} = \tfrac{2}{3} \epsilon_{20}
    + \tfrac{5}{63} \epsilon_{20}^2
    - \tfrac{2}{21} \epsilon_{20}\epsilon_4
    - \tfrac{10}{63} \epsilon_{22}^2
    + \tfrac{50}{231} \epsilon_4^2 , \\
  \alpha_{22} = \alpha_{2(-2)} = \tfrac{2}{3} \epsilon_{22}
    - \tfrac{10}{63} \epsilon_{20} \epsilon_{22}
    - \tfrac{1}{63} \epsilon_{22} \epsilon_4 , \\
  \alpha_{40} = - \epsilon_4
    + \tfrac{12}{35} \epsilon_{20}^2
    - \tfrac{30}{77} \epsilon_{20}\epsilon_4
    + \tfrac{4}{35} \epsilon_{22}^2
    + \tfrac{243}{1001} \epsilon_4^2 , \\
  \alpha_{42} = \alpha_{4(-2)}
    = \sqrt{\tfrac{48}{245}} \epsilon_{20} \epsilon_{22}
    + \sqrt{\tfrac{1215}{5929}} \epsilon_{22} \epsilon_4 , \\
  \alpha_{44} = \alpha_{4(-4)}
    = \sqrt{\tfrac{8}{35}} \epsilon_{22}^2 .
\end{gathered}\end{equation}
This approximation is adopted. (For $\epsilon_{22} = 0$ the
expansion~\eqref{eq:ep>al} (including results for $l > 4$ which we do
not show) should give Eqs.~(10)--(13) of Ref.~\cite{ref:See75}. Some
coefficients there differ from ours, which were derived by computer
algebra.)

The coefficients with $l > 4$ are not required in the second step,
where $B_s$ and $B_c$ are expanded in the $\alpha$'s. This expansion
can be derived from Swiatecki's results in Ref.~\cite{ref:Swi56}.
Swiatecki's expansion is restricted to $\gamma = 0$, but when only
terms of total rank 8 or less are retained, each term has a unique
continuation into $\gamma \ne 0$ given by the requirement that it be a
scalar polynomial in the spherical tensor components $\alpha_{lm}$.
The resulting expansion, which we adopt, is
\begin{equation}\begin{gathered}
  B_s = 1 + \tfrac{2}{5}p_{20} - \tfrac{4}{105}p_{30} 
    - \tfrac{66}{175}p_{40} - \tfrac{4}{35}p_{21} + p_{02},\\
  B_c = 1 - \tfrac{1}{5}p_{20} - \tfrac{4}{105}p_{30}
    + \tfrac{51}{245}p_{40} - \tfrac{6}{35}p_{21} - \tfrac{5}{27}p_{02}
\end{gathered}\end{equation}
with
\begin{equation}\begin{gathered}
  p_{20} = \alpha_{20}^2 + 2\alpha_{22}^2 , \\
  p_{30} = \alpha_{20}(\alpha_{20}^2 - 6\alpha_{22}^2) , \\
  p_{40} = p_{20}^2 , \\
  p_{21} = (\alpha_{20}^2 + \tfrac{1}{3}\alpha_{22}^2)\alpha_{40}
    + \sqrt{\tfrac{20}{3}}\alpha_{20}\alpha_{22}\alpha_{42}
    + \sqrt{\tfrac{70}{9}}\alpha_{22}^2\alpha_{44} , \\
  p_{02} = \alpha_{40}^2 + 2\alpha_{42}^2 + 2\alpha_{44}^2 .
\end{gathered}\end{equation}

For given pairing parameters $\bar G$, $\zeta$, $\alpha$ and an RPA
interpolation width $w$ defined in Sec.~\ref{sec:int} we fix the
coefficients $a_x$ in Eq.~\eqref{eq:LD} by a least-square fit of the
calculated total energies~\eqref{eq:E} to the measured ones. Included
in this fit are all doubly even nuclei in the considered region of the
chart of nuclei whose binding energies have been measured. The limits
of each region for this purpose are specified in
Secs.~\ref{sec:N=Z}--\ref{sec:Zr}. The fit of the liquid drop
parameters $a_x$ is done before the pairing parameters are fit to
other data. Table~\ref{tbl:LD} shows the results for the optimal
pairing parameters. For the $^{102}$Zr region the sample of doubly
even nuclei consists of only 9 nuclei.
\begin{table}
  \caption{\label{tbl:LD}Liquid drop parameters for optimal pairing
    parameters. The last column shows the rms deviation from the data.
    The unit is MeV  throughout.}
\begin{ruledtabular}
\begin{tabular}{lcccccc}
&$a_v$&$a_{vt}$&$a_s$&$a_{st}$&$a_c$&rms\\
\hline
$N \approx Z$&15.23&112.5&16.52&148.9&0.6601&1.018\rule{0pt}{1em}\\
Around Sn&15.37&115.2&16.97&157.5&0.6737&0.515\\
Around $^{102}$Zr&14.78&151.2&16.07&355.5&0.5774&0.043
\end{tabular}
\end{ruledtabular}
\end{table}

\subsection{\label{sec:in}Independent nucleons}

The terms $E_\text{i.n.,$\tau$}$ in Eq.~\eqref{eq:mic} are given by
\begin{equation}
  E_\text{i.n.,$\tau$} = \sum_{k=1}^{N_\tau} \epsilon_{k \tau} ,
\end{equation}
with $N_\tau = N$ for $\tau = n$ and $N_\tau = Z$ for $\tau = p$. The
single-nucleon energies $\epsilon_{k \tau}$ are the eigenvalues of the
Nilsson Hamiltonian~\cite{ref:Nil55,ref:Nil69,ref:Lar73},
\begin{multline}\label{eq:hNil}
  h_\tau = \frac{\bm p^2}{2 M_\tau} + \tfrac12 \left(
    M_\tau \sum_{q = 1}^3 (\omega_\alpha x_\alpha)^2
    + 2 \epsilon_4 \omega_0 \rho^2
      P_4(\cos \theta_{\text t}) \right) \\
    - \kappa_{N_\text{sh},\tau} \overset\circ\omega
      \left( 2 \bm l_{\text t} \cdot \bm{s}
        + \mu_{N_\text{sh},\tau} ( \bm l_{\text t}^2
          - \langle \bm l_{\text t}^2 \rangle_{N_\text{sh}} )
      \right) ,
\end{multline}
where $\bm r = (x_1,x_2,x_3)$ and $\bm p$ are the spatial coordinates
and momentum, $\bm s$ is the spin, and $M_\tau$ is the nucleon mass.
The function $P_l(x)$ is the Legendre polynomial. The oscillator
frequencies $\omega_q$ are given by
\begin{equation}
  \omega_q = \omega_0 \left( 1 
    - \tfrac23 \epsilon_2 \cos (\gamma + q \tfrac{2\pi}3) \right) ,
\end{equation}
where $\omega_0$ satisfies  the condition of volume conservation

\begin{equation}
  \prod_{q = 1}^3 \omega_q = \overset\circ\omega^3 , \quad
  \overset\circ\omega = 41 A^{-1/3}~\text{MeV} .
\end{equation}
The 'stretched' spherical coordinates
$(\rho,\theta_{\text t},\phi_{\text t})$ and orbital angular momentum
$\bm l_{\text t}$~\cite{ref:Nil55} correspond to Cartesian coordinates
\begin{equation}
  \xi_{q\tau} = x_q \sqrt {M_\tau \omega_q} ,
\end{equation}
and $N_\text{sh}$ is the number of oscillator quanta. For the
parameters $\kappa_{N_\text{sh},\tau}$ and $\mu_{N_\text{sh},\tau}$ we
adopt the values recommended in Ref.~\cite{ref:Ben85}.

The independent-nucleon counter terms are
\begin{equation}
  \tilde E_\text{i.n.,$\tau$}
    = 2 \int\limits_{-\infty}^{\tilde \lambda_\tau}
      \epsilon \tilde g_\tau(\epsilon) d \epsilon ,
\end{equation}
where the smooth chemical potential $\tilde \lambda_\tau$ is defined
by
\begin{equation}\label{eq:smlam}
  2 \int\limits_{-\infty}^{\tilde \lambda_\tau}
      \tilde g_\tau(\epsilon) d \epsilon = N_\tau
\end{equation}
and the smooth level density $\tilde g_\tau(\epsilon)$ is given
by~\cite{ref:Str66,ref:Rin80}
\begin{multline}\label{eq:smg}
   \tilde g_\tau(\epsilon) = \frac1{2 \gamma_\text{Str} \sqrt{\pi}}
   \sum_k L \left( m_\text{Str} , \tfrac12 , \left( \frac
       {\epsilon - \epsilon_{k \tau}} {\gamma_\text{Str}}
     \right)^2 \right) \\
     \exp \left( - \left( \frac
       {\epsilon - \epsilon_{k \tau}} {\gamma_\text{Str}}
     \right)^2 \right)
\end{multline}
in terms of the generalized Laguerre polynomial $L(n,a,x)$. We use
smoothing width $\gamma_\text{Str} = \overset\circ\omega$ and
smoothing order $m_\text{Str} = 3$ and include in the sum in
Eq.~\eqref{eq:smg} all such $k$ that
$\epsilon_{k \tau} < 47.5~\text{MeV} + 5 \, \gamma_\text{Str}$ and
$N_\text{sh} \le 9$.

\subsection{\label{sec:BCS}BCS theory}

The terms $E_\text{BCS,$\tau$}$ are given by the standard BCS theory.
A derivation of the following equations is found, for example in
Ref.~\cite{ref:Nee09}. For even $N_\tau$ one has
\begin{equation}
  E_\text{BCS,$\tau$}
    = \sum_{k=1}^{2 \Omega_\tau} v_{k \tau}^2 \epsilon_{k \tau}
      - \frac {\Delta_\tau^2} {G_\tau}
      - E_\text{i.n.,$\tau$}
\end{equation}
with
\begin{equation}\label{eq:uv}\begin{gathered}
  \left. \begin{matrix} u_{k \tau} \\ v_{k \tau} \end{matrix} \right\}
    = \sqrt{ \frac12 \left( 1 \pm \frac
      {\epsilon_{k \tau} - \lambda_\tau} {E_{k \tau}} \right) } , \\
  E_{k \tau} = \sqrt{(\epsilon_{k \tau} - \lambda_\tau)^2
    + \Delta_\tau^2} .
\end{gathered}\end{equation}
Here $\lambda_\tau$ and $\Delta_\tau$ obey
\begin{equation}~\label{eq:BCS}
  \sum_{k=1}^{2 \Omega_\tau} v_{k \tau}^2 = N_\tau , \quad
  G_\tau \sum_{k=1}^{2 \Omega_\tau} u_{k \tau}  v_{k \tau}
     = 2 \Delta_\tau .
\end{equation}
For later reference we define the quasinucleon annihilators
\begin{equation}
  \alpha_{k\tau} =u_{k\tau} a_{k\tau}
    - v_{k\tau} a_{\bar{k\tau}}^\dagger .
\end{equation}

The equations~\eqref{eq:uv} and \eqref{eq:BCS} always have a
solution with $\Delta_\tau = 0$ and there is a threshold
$G_\text{cr,$\tau$}$ such that no other $\Delta_\tau$ is possible for
$G \le G_\text{cr,$\tau$}$. For $G > G_\text{cr,$\tau$}$ there is a
solution with $\Delta_\tau > 0$ and a lower $E_\text{BCS,$\tau$}$,
which is chosen. If
$\epsilon_{(N_\tau+2) \tau} > \epsilon_{N_\tau \tau}$ then
$G_\text{cr,$\tau$} > 0$ and $G_\text{cr,$\tau$}$ is given by
\begin{equation}~\label{eq:Gcr}
  \frac 4 {G_\text{cr,$\tau$}} = \min_{\epsilon_{N_\tau \tau} <
    \lambda_\tau < \epsilon_{(N_\tau+2) \tau}}
  \sum_{k=1}^{2 \Omega_\tau} \frac 1 {|\epsilon_{k \tau} -
    \lambda_\tau|} .
\end{equation}
If $\epsilon_{(N_\tau+2) \tau} = \epsilon_{N_\tau \tau}$, as happens
in spherical nuclei when a $j$ shell is partly occupied in the absence
of pairing, then $G_\text{cr,$\tau$} = 0$.

If $N_\tau$ is odd, a Bogolyubov quasinucleon annihilated by
$\alpha_{N_\tau \tau}$ is assumed to be
present in the BCS ground state. The orbit $|N_\tau \tau \rangle$ is
then fully occupied and its time reverse $|(N_\tau + 1) \tau \rangle$
fully empty. The BCS energy $E_\text{BCS,$\tau$}$ is calculated as if
$N_\tau -1$ nucleons of type $\tau$ inhabited the remaining orbits.
The odd nucleon is said to \textit{block the Fermi level}.

To simplify notation we let $\tilde g_\tau$ without an argument mean
$\tilde g_\tau(\tilde \lambda_\tau)$ and write
\begin{equation}\label{eq:chi}
  \frac 1 {\tilde g_\tau G_\tau} = \chi_\tau .
\end{equation}
The BCS counter terms are then given by~\cite{ref:Str67,ref:Nee16}
\begin{equation}\label{eq:smBCS}
  \tilde E_\text{BCS,$\tau$}
    = - \tfrac12 \Omega_\tau \tilde \Delta_\tau \exp (- \chi_\tau) ,
  \quad
  \tilde \Delta_\tau = \frac {\Omega_\tau}
    {2 \tilde g_\tau \sinh \chi_\tau} .
\end{equation}

\subsection{\label{sec:RPA}Random-phase approximation}

The calculation of $E_\text{RPA,$\tau$}$ is based on the theory in
Ref.~\cite{ref:Nee09}. It involves linear relations in the space
spanned by the terms in the sums in Eq.~\eqref{eq:P}. A linearly
independent set of terms in the expression for $P_\tau$ may be labeled
by the odd single-nucleon indices $k$ from 1 to $2 \Omega_\tau - 1$.
When both $N$ and $Z$ are even, we denote this set of $k$ by
$\mathcal S_\tau$. Modifications of this definition when one or both
of $N$ and $Z$ are odd are discussed below. It is convenient to
introduce at this point labels $\tau\tau' = nn, pp, np$ alternative to
and synonymous with $\tau = n, p, np$ and vectors and matrices
with components or element indexed by the set
$\mathcal S_{\tau\tau'}$. A diagonal matrix $\mathsf{E_{\tau\tau'}}$
is defined by its elements
\begin{equation}
  E_{\tau\tau',kl} =
     \delta_{kl} (E_{k \tau} + E_{k \tau'})
\end{equation}
and column vectors $\mathsf{U_{\tau\tau'}}$ and
$\mathsf{V_{\tau\tau'}}$ by their components
\begin{equation}\label{eq:UV}
  U_{\tau\tau',k} = u_{k \tau} u_{k \tau'}  , \quad
  V_{\tau\tau',k} = - v_{k \tau} v_{k\tau'} .
\end{equation}
Let
\begin{equation}\begin{gathered}
  \mathsf{A_{\tau\tau'}} = \mathsf{E_{\tau\tau'}}
    - G_{\tau\tau'} \left(
      \mathsf{U_{\tau\tau'}} \mathsf{U_{\tau\tau'}}^T
      + \mathsf{V_{\tau\tau'}} \mathsf{V_{\tau\tau'}}^T \right) , \\
  \mathsf{B_{\tau\tau'}} = - G_{\tau\tau'} \left(
      \mathsf{U_{\tau\tau'}} \mathsf{V_{\tau\tau'}}^T
      + \mathsf{V_{\tau\tau'}} \mathsf{U_{\tau\tau'}}^T \right) .
\end{gathered}\end{equation}
Then 
\begin{equation}\label{eq:ERPA}
  E_\text{RPA,$\tau\tau'$} = \tfrac12 \left(
    \sum_k \sqrt{z_{\tau\tau',k}}
    - \text{tr} \, \mathsf{E_{\tau\tau'}} \right) ,
\end{equation}
where $z_{\tau\tau',k}$ are the eigenvalues of
\begin{equation}
   (\mathsf{A_{\tau\tau'}} + \mathsf{B_{\tau\tau'}})
   (\mathsf{A_{\tau\tau'}} - \mathsf{B_{\tau\tau'}}) .
\end{equation}
The terms $\sqrt{z_{\tau\tau',k}}$ are the RPA frequencies.

For $\tau = \tau'$ and, in the limit of isobaric invariance, for
$\tau\tau' = np$ and $N = Z$, one RPA mode is, for
$G_{\tau\tau'} > G_\text{cr,$\tau\tau'$}$ (with $G_\text{cr,$np$} =
G_\text{cr,$n$} = G_\text{cr,$p$}$ in the isobarically invariant
limit), a Nambu-Goldstone mode with zero
frequency~\cite{ref:Nee09,ref:Hin16}. That is, in this degree of
freedom vibration turns into rotation. This is what gives rise to the
singularity at $G = G_\text{cr}$ in
Fig.~\ref{fig:bang}~\cite{ref:Ben14}. To circumvent this singularity
we interpolate the calculated $E_\text{RPA,$\tau\tau'$}$ across the
region of $G_{\tau\tau'} = G_\text{cr,$\tau\tau'$}$ for $\tau = \tau'$
or $\tau\tau' = np$ and $N = Z$ with $G_\text{cr,$np$} \approx
G_\text{cr,$n$} \approx G_\text{cr,$p$}$ in the latter case. Details
are given in Sec.~\ref{sec:int}.

The expression~\eqref{eq:ERPA} results from the expansion of the
ground state energy in Feynman diagrams formed as closed bubble
chains; see Eq.~(36) in Ref.~\cite{ref:Nee09}. Each bubble represents
a virtual creation and subsequent annihilation of a pair of Bogulyubov
quasinucleons. When, say, $N$ is odd, the presence of the unpaired
nucleon in the BCS ground state blocks the creation of quasinucleon
pairs by the terms in $P_n$ and $P_n^\dagger$ proportional to
$\alpha_{N n}^\dagger \alpha_{(N+1) n}^\dagger$. Therefore $k = N$
should be and is omitted from $\mathcal S_n$ for odd $N$. The
remainder exhausts the set of excitations of the BCS ground state
mediated by the fields $P_n$ and $P_n^\dagger$.

The case of $\mathcal S_{np}$ is more involved for odd $N$. The fields
$P_{np}$ and $P_{np}^\dagger$ have terms proportional to
$\alpha_{(N+1) n}^\dagger \alpha_{N p}^\dagger$ and
$\alpha_{N p}^\dagger \alpha_{N n}$, which, respectively, adds a pair
of quasinucleons and scatters the quasineutron in the Fermi level
orbit into a quasiproton. The latter excitation, in particular, may
have negative energy, which inhibits the use of the RPA. Even when the
energy is positive, it is small in comparison to that of the genuine
two-quasinucleon excitations, which may render the RPA calculation
unstable anyway. For $Z = N$ in the limit of isobaric invariance, both
these excitations have zero matrix elements when one assumes, as we
do, cf. Sec.~\ref{sec:alog}, that the unpaired neutron and the
unpaired proton combine to isospin $T = 0$. This allows using
Eq.~\eqref{eq:ERPA}, omitting $k = N$ from $\mathcal S_{np}$ like it
is omitted from $\mathcal S_n$. To avoid the troubles just described,
we have chosen to do so also when $Z$ is even. That is, we generally
omit $k = N$ from $\mathcal S_n$ and $\mathcal S_{np}$ when $N$ is
odd, and analogously for odd $Z$. In physical terms this amounts to
extending to the RPA the assumption in the BCS theory with the Fermi
level blocked that the unpaired nucleon acts as a spectator to
interactions among the paired nucleons in a valence space that
excludes the half occupied single-nucleon level. A more satisfactory
treatment of the neutron-proton pair vibrational correlations for odd
$A$ might be based on the theory of (quasi-) \linebreak
particle-vibration coupling.

For even $N$ and $Z$ the RPA energy as given by Eq.~\eqref{eq:ERPA}
gets contributions from fluctuations of the quasinucleon vacuum in
every direction generated by an operator
$\alpha_{k\tau}^\dagger \alpha_{\bar{k\tau'}}^\dagger +
\alpha_{k\tau'}^\dagger \alpha_{\bar{k\tau}}^\dagger$ with
$k \in \mathcal S_{\tau\tau'}$. Vaquero, Egido, and Rodr\'iguez take
an different path to study pairing fluctuations~\cite{ref:Vac13}. A
combination of the variances of $N$ and $Z$ is used (for a given
deformation) as a generator coordinate to obtain a wave function that
describes the distribution of quasinucleon vacua in the single degree
of freedom associated with this coordinate. The quasinucleon vacua are
generated by the constrained Hartree-Fock-Bogolyubov method with a
Gogny two-nucleon interaction.

For the calculation of the RPA counter terms
$\tilde E_\text{RPA,$\tau\tau'$}$ we define
$\tilde g_{\tau\tau'}(\epsilon)$ by replacing $\epsilon_{k\tau}$ by
$(\epsilon_{k\tau} + \epsilon_{k\tau'})/2$ in the expression~\eqref{eq:smg}.
This definition coincides with Eq.~\eqref{eq:smg} for $\tau=\tau'$. A
function $\tilde \lambda_{\tau\tau'}(x)$ is defined by
\begin{equation}
  2 \int\limits_{-\infty}^{\tilde \lambda_{\tau\tau'}(x)}
      \tilde g_{\tau\tau'}(\epsilon) d \epsilon = x .
\end{equation}
In particular $\tilde \lambda_{\tau\tau}(N_\tau) = \tilde\lambda_\tau$
by Eq.~\eqref{eq:smlam}. We let $\tilde g_{np}$ without an
argument mean $\tilde g_{np}(\tilde \lambda_{np} (A/2))$ and
generalize Eq.~\eqref{eq:chi} to
\begin{equation}\label{eq:chinp}
  \frac 1 {\tilde g_{\tau\tau'} G_{\tau\tau'}} = \chi_{\tau\tau'}
\end{equation}
and the definition of $\tilde \Delta_\tau$ in Eq.~\eqref{eq:smBCS} to
\begin{multline}
  \tilde \Delta_{\tau\tau'} = \frac {\Omega_{\tau\tau'}}
      {2 \tilde g_{\tau\tau'} \sinh \chi_{\tau\tau'}}
    \\ \sqrt {1 -
      \left( \frac {\tilde g_{\tau\tau'} (
          \tilde \lambda_{\tau\tau'}(N_\tau)
      - \tilde \lambda_{\tau\tau'}(N_{\tau'}) )
          \tanh \chi_{\tau\tau'} } {\Omega_{\tau\tau'}} \right)^2 } .
\end{multline}
Then $\tilde E_\text{RPA,$\tau\tau'$}$ is given by~\cite{ref:Nee16}
\begin{multline}\label{eq:smRPA}
  \tilde E_\text{RPA,$\tau\tau'$}
  = \frac {2 \tilde \Delta_{\tau\tau'}} \pi
    \int_0^\infty \ln \Biggl( \frac 1 {\chi_{\tau\tau'}}  \\
      \tanh^{-1} \left(
        \left( 1 + (l_{\tau\tau'}^2 + x^2)^{-1} \right)^{-\frac12}
      \tanh \chi_{\tau\tau'} \right) \Biggr) dx
\end{multline}
with
\begin{equation}\label{eq:l}
  l_{\tau\tau'} = \frac {\tilde \lambda_{\tau\tau'}(N_\tau)
      - \tilde \lambda_{\tau\tau'}(N_{\tau'})}
    {2 \tilde \Delta_{\tau\tau'}} .
\end{equation}

\subsection{\label{sec:alog}Isobaric  analogs}

The scheme presented so far describes states with isospin
$T \approx |M_T|$. This relation is satisfied empirically by nearly
all ground states. The exception is that for odd $N = Z > 20$ most
ground states have $T \approx 1$ while the lowest states with
$T \approx 0$ are excited. For odd $N = Z < 20$ the lowest states with
$T \approx 1$ are mostly excited. We denote the energies of these
$T \approx 1$ states by $E^\ast(N,Z)$ to distinguish them from the
energies of the $T \approx 0$ states. For odd $N = Z$ the
$T \approx 1$ states are the isobaric analogs of the ground states of
the doubly even nuclei with neutron and proton numbers
$(N',Z') = (N + 1,Z - 1)$. Accordingly we set
\begin{multline}
  E^*(N,Z) = E(N',Z') \\
    + a_c \frac{Z(Z-1) - Z'(Z'-1)}{A^{1/3}} B_c ,
\end{multline}
where $B_c$ is calculated from the deformation of the doubly even
nucleus.

\section{\label{sec:N=Z}$N \approx Z$ region}

Our calculations for even $A$ in the $N \approx Z$ region follow the
scheme previously applied in Refs.~\cite{ref:Ben14,ref:Nee17}. Again
we consider the doubly even nuclei with $24 \le A \le 100$ and
$0 \le N - Z \le 10$ and the doubly odd ones with $26 \le A \le 98$
and $N = Z$. Unlike Ref.~\cite{ref:Nee17} we use different
$\Omega_\tau$ for different $\tau$ and a considerably smaller interval
of interpolation of the RPA energies as discussed in
Sec.~\ref{sec:int}. Further, the deformations were recalculated, all
oscillator shells with $N_\text{sh} \le 9$ being included in the
calculation by the scheme of Ref.~\cite{ref:Ben10} instead of just
four shells close to the neutron or proton Fermi level for $\tau = n$
and $p$, respectively. For the doubly even nuclei this only changed
the deformations of $^{84}$Zr and $^{86}$Mo, which went from spherical
to oblate. For the $T \approx 0$ states of the doubly odd nuclei, the
deformations were determined in the prior work by averaging over the
deformations of the adjacent doubly even nuclei. In the present work
these deformations are calculated independently by blocking the Fermi
levels. This resulted in significant changes of the individual
deformations, while the overall pattern of variation along the chain
of these states remains the same.

Again we set $\alpha = 0$ in Eq.~\eqref{eq:G} so that one pair
coupling constant $G$ covers the cases $\tau = n$, $p$, and $np$. The
parameters $\bar G$ and $\zeta$ are fit to the following data for odd
$N = Z$.

(1) The $T \approx 0$ doubly odd--doubly even mass differences
\begin{equation}
   E(N,N) -\tfrac12 [ E(N\!-\!1,N\!-\!1) + E(N\!+\!1,N\!+\!1) ] .
\end{equation}

(2) The differences of the lowest energies for $T \approx 1$ and
$T \approx 0$, that is,
\begin{equation}
   E^*(N,N) - E(N,N) .
\end{equation}
The set of data is the same as in Refs.~\cite{ref:Ben14,ref:Nee17} and
thus includes extrapolated masses of $^{82}$Nb and $^{86}$Tc, but all
mass data were updated from AME12~\cite{ref:Aud12} to
AME16~\cite{ref:Hua17}. Again excitation energies are taken from the
Evaluated Nuclear Structure Data File~\cite{ref:ENSDF}. A least-square
fit gives
\begin{equation}\label{eq:GN=Z}
  G = 7.196 A^{-0.7461} \text{ MeV}
\end{equation}
with an rms deviation of 0.789~MeV. Plotting the $T \approx 0$ doubly
odd--doubly even mass differences, the $T \approx 0$ to $T \approx 1$
energy splittings, the symmetry energy coefficients, and the `Wigner
$x$' as functions of $A$ results in figures grossly similar to Figs.
6--9 of Refs.~\cite{ref:Ben14} and Fig. 1 of Ref.~\cite{ref:Nee17}. As
for the Wigner $x$, more detail is given in Sec.~\ref{sec:int}.

\begin{figure*}
  {\centering\includegraphics{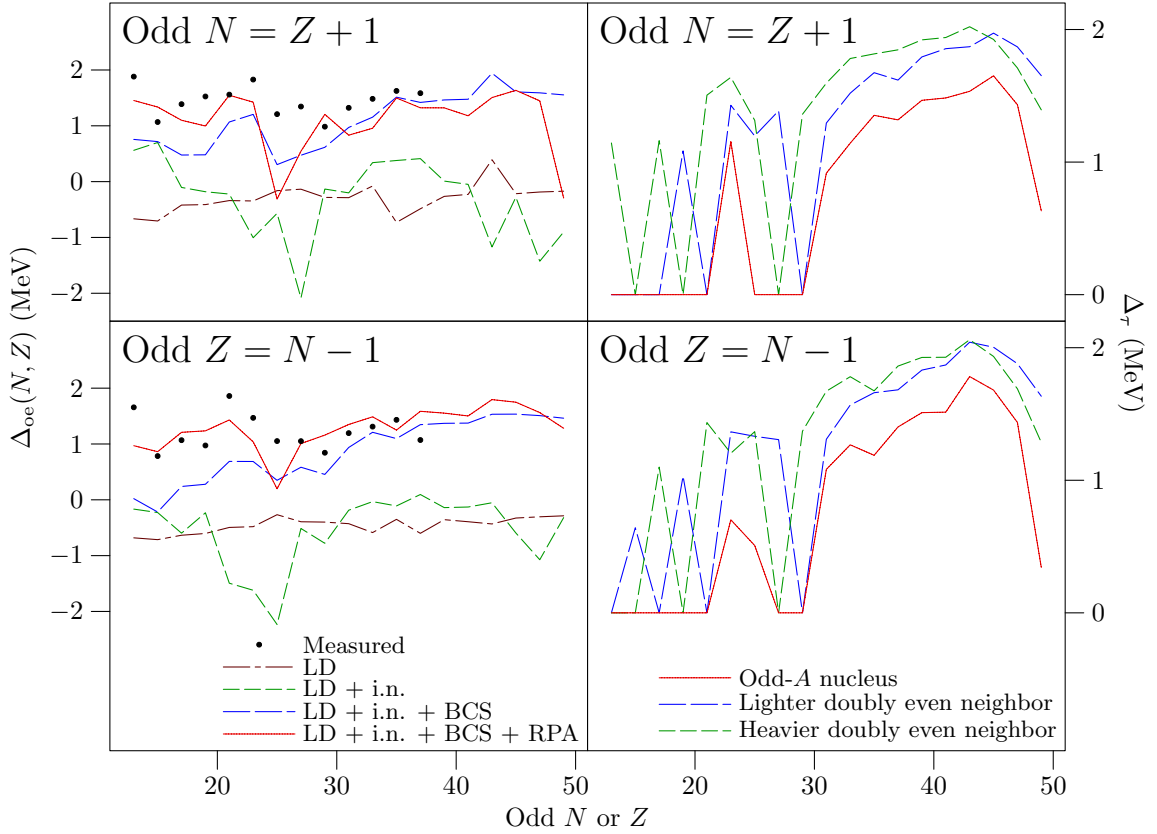}\par}
  \caption{\label{fig:oeN=Z}(Color online)  The panels on the left
    show the calculated odd-even mass differences
    $\Delta_\text{oe}(N,Z)$ for $Z = N - 1$ in successive
    approximations in comparison with the values extracted from mass
    data. The panels on the right display the BCS gap parameters
    $\Delta_\tau$ of the odd-$A$ nuclei and their doubly even
    neighbors, where $\tau = n$ for odd $N$ and $p$ for odd $Z$.  }
\end{figure*}
With the parameters thus set we consider the odd-$A$ nuclei with
$Z = N - 1$ and $25 \le A \le 99$. The odd-even mass difference
$\Delta_\text{oe}(N,Z)$ is defined as the mass of the odd-$A$ nucleus
relative to the average mass its two doubly even neighbors. The
calculated $\Delta_\text{oe}(N,Z)$ are shown in Fig.~\ref{fig:oeN=Z}
in comparison with the data. The model is seen to reproduce the
typical size of the measured values. This is remarkable because
$\bar G$ and $\zeta$ were fit, not to these data but to \emph{energies
  in doubly odd nuclei}. This supports an interpretation of the lowest
$T \approx 0$ states of such nuclei as essentially two-quasinucleon
states.

The figure also displays the individual contributions to the
calculated $\Delta_\text{oe}(N,Z)$ from $E_\text{LD}$, \linebreak
$\delta E_\text{i.n.} = \sum_{\tau = n,p} \delta
E_\text{i.n.,$\tau$}$,
$\delta E_\text{BCS} = \sum_{\tau = n,p} \delta E_\text{BCS,$\tau$}$,
and
$\delta E_\text{RPA} =
\sum_{\tau = n,p,np} \delta E_\text{RPA,$\tau$}$. The liquid drop
contribution is negative except for $N = 43$ with an average
about~$-0.4$~MeV. The contribution from the independent-nucleon shell
correction $\delta E_\text{i.n.}$ fluctuates wildly as a function of
$N$ or $Z$. These fluctuations are reduced by the pairing, which also
renders the total $\Delta_\text{oe}(N,Z)$ mostly positive in
accordance with the data. Very low and, for odd $N$, even negative
values are calculated, however, for $N$ and $Z = 25$ and for $N = 49$,
not the least induced by anomalously low contributions of $\delta
E_\text{RPA}$. These low contributions, as well as one at $Z = 49$,
are correlated with $G_\text{cr,$n$}$ or $G_\text{cr,$p$}$ being close
to $G$ for odd $N$ and $Z$, respectively, so that the accuracy of the
RPA is uncertain, cf.~Sec.~\ref{sec:int}. The measured odd-even mass
difference actually decreases when $N$ or $Z = 25$ is approached from
below, but this decrease is much exaggerated in the calculation.

\begin{figure}
  \centering\includegraphics{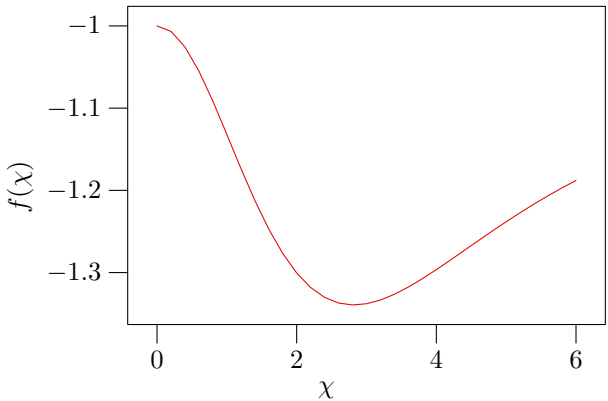}
  \caption{\label{fig:f}(Color online) The function $f$  given by
    Eq.~\eqref{eq:f}.}
\end{figure}
The RPA contribution is positive for all odd $Z$ except $Z = 25$ and
49 and for all odd $N < 30$ except $N = 25$. In the upper \textit{sd}
shell it gives almost the entire $\Delta_\text{oe}(N,Z)$ for odd $Z$
and about half of it for odd $N$. For odd $N > 30$ the RPA
contribution is negative, and both for odd $N$ and for odd $Z$ it is
numerically smaller in the heavier than in the lighter nuclei.

These differences in the size and sign of the RPA contribution may be
understood qualitatively from the expression~\eqref{eq:smRPA}. Thus
for $l_{\tau\tau'} = 0$, which holds by Eq.~\eqref{eq:l} for
$\tau = \tau'$ and approximately for $\tau\tau' = np$ and
$N \approx Z$, Eqs.~\eqref{eq:chinp}--\eqref{eq:smRPA} give
\begin{equation}
  \tilde E_\text{RPA,$\tau\tau'$}
    = \tfrac12 \, \Omega_{\tau\tau'} G_{\tau\tau'} f(\chi_{\tau\tau'})
\end{equation}
with
\begin{multline}\label{eq:f}
  f(\chi) = \frac {2 \chi} {\pi \sinh \chi}
    \int_0^\infty \ln \Biggl( \frac 1 {\chi} \\
      \tanh^{-1} \left( \left( 1 +  x^{-2} \right)^{-\frac12}
      \tanh \chi \right) \Biggr) dx .
\end{multline}
This function is displayed in Fig.~\ref{fig:f}. The contribution of
$\delta E_\text{RPA}$ to $\Delta_\text{oe}(N,Z)$ stems mainly from the
microscopic term $E_\text{RPA}$. In fact, because the counter term
$\tilde E_\text{RPA}$ is a smooth function of $N$, $Z$, and
deformation, with no distinction between even and odd $N_\tau$, its
contribution is small. Consider the case of odd $N$. The difference
between $E_\text{RPA,$n\tau$}$ for odd and even $N$ is roughly a
result of the effective dilution in the odd case of the single-neutron
spectrum by the blocking of the Fermi level. The impact on
$E_\text{RPA,$n\tau$}$ of this decrease of level density near the
Fermi level is similar to the impact on $\tilde E_\text{RPA,$n\tau$}$
of a decrease of $\tilde g_{n\tau}$. By Eqs.~\eqref{eq:chi}
and~\eqref{eq:chinp} the latter increases $\chi_{n\tau}$ and thus
gives rise to an increase of $\tilde E_\text{RPA,$n\tau$}$
proportional to $f'(\chi_{n\tau})$ with a positive coefficient. The
case of odd $Z$ is analogous. The calculated $\chi_{\tau\tau'}$
decrease from about 3.8 for $A = 24$ to about 2.6 for $A = 100$. Thus
in the lighter nuclei we have $f'(\chi_{\tau\tau'}) > 0$ and
accordingly expect a large positive RPA contribution to
$\Delta_\text{oe}(N,Z)$, while in the heavier nuclei we have
$f'(\chi_{\tau\tau'}) \approx 0$ and accordingly expect a small
contribution, which can take either sign.

Also shown in Fig~\ref{fig:oeN=Z} are the calculated gap
parameters $\Delta_\tau$ for both the odd-$A$ nucleus and its doubly
even neighbors. It is seen that often in the lighter nuclei,
$\Delta_\tau = 0$, most often for odd $A$. The BCS approximation to
$\Delta_\text{oe}(N,Z)$ is seen to follow roughly the fluctuating gap
parameters as a function of $N$ or $Z$.

\section{\label{sec:Sn}Neighborhood of the $\mathrm{Sn}$ isotopes}

In the neighborhood of the Sn isotopic chain we consider all nuclei
with $48 \le Z \le 52$ and even $N$ in the interval $46 \le N \le 92$
and all Sn isotopes with odd $N$ in the interval $47 \le N \le 91$. In
Eq.~\eqref{eq:G}, we keep the $A$ exponent $\zeta = -0.7461$ which
resulted from the analysis of data for $N = Z$, cf.
Eq.~\eqref{eq:GN=Z}, but adjust $\bar G$ and $\alpha$ so as to
reproduce the average of the measured $\Delta_\text{oe}(N,Z)$
separately for odd $N$ and odd $Z$. The result is
\begin{equation}\label{eq:GSn}
  G_\tau = 5.818 A^{-0.7461} (1 - 0.0170 M_T M_T') \text{ MeV} .
\end{equation}
For $^{100}$Sn, Eq.~\eqref{eq:GN=Z} gives $G_\tau = 0.2317$~MeV for
all $\tau$, while Eq.~\eqref{eq:GSn} gives $G_\tau = 0.1873$~MeV for
all $\tau$. We thus have two determinations of the pair coupling
constant in $^{100}$Sn, the higher one 24\% greater than the lower
one. They result from extrapolation from different directions in the
chart of nuclei, one from the $N = Z$ line and one from the
neighborhood of the Sn isotopic chain. Because the data in the
fit~\eqref{eq:GN=Z} include extrapolated masses and interpretations of
incomplete spectra of $^{82}$Nb and $^{86}$Tc, the lower value is
likely to be most reliable.

\begin{figure*}
  {\centering\includegraphics{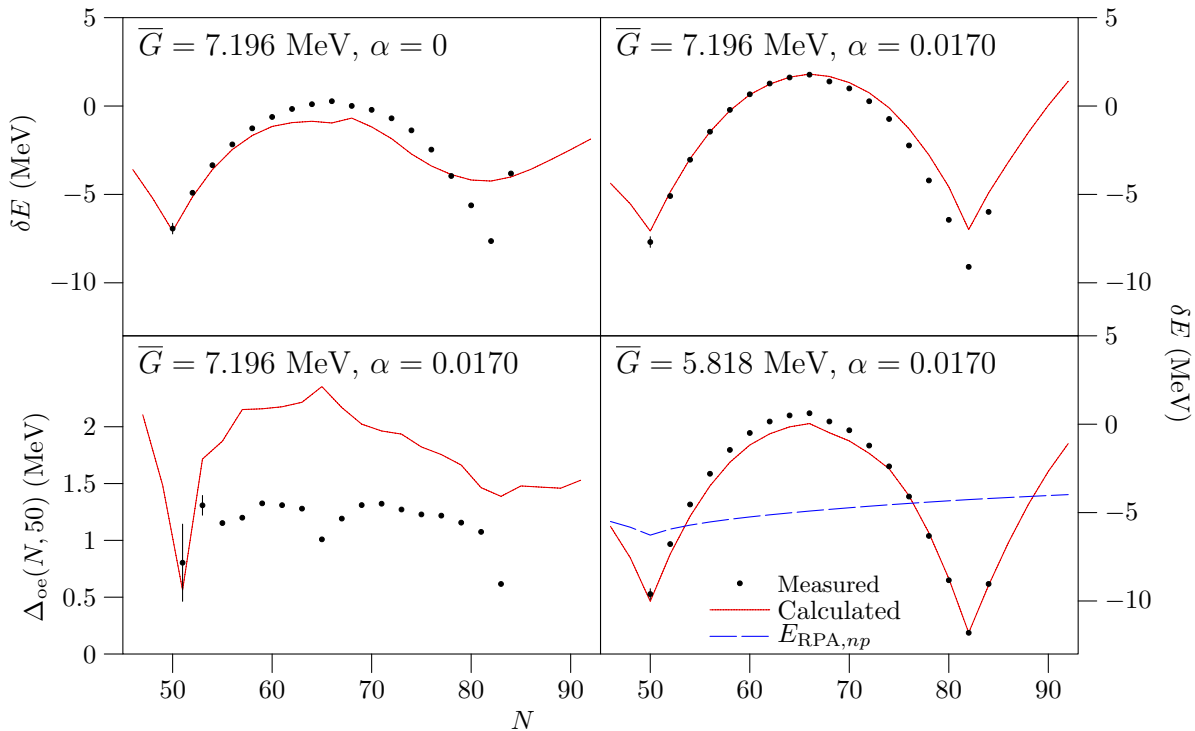}\par}
  \caption{\label{fig:GSn}(Color online) The upper left, upper right,
    and lower right panels show the calculated shell corrections
    $\delta E$ of the doubly even Sn isotopes for three different sets
    of pairing parameters. The lower left panel shows the odd-even
    mass differences $\Delta_\text{oe}(N,50)$ calculated with the
    pairing parameters of the upper right panel. For all these results
    the corresponding empirical values are shown for comparison. The
    empirical shell corrections $\delta E_\text{emp}$ differ between
    the panels due to different liquid drop parameters. A plot of the
    neutron-proton RPA energy $E_\text{RPA,$np$}$ is included in the
    lower right panel.}
\end{figure*}
\begin{figure*}
  {\centering\includegraphics{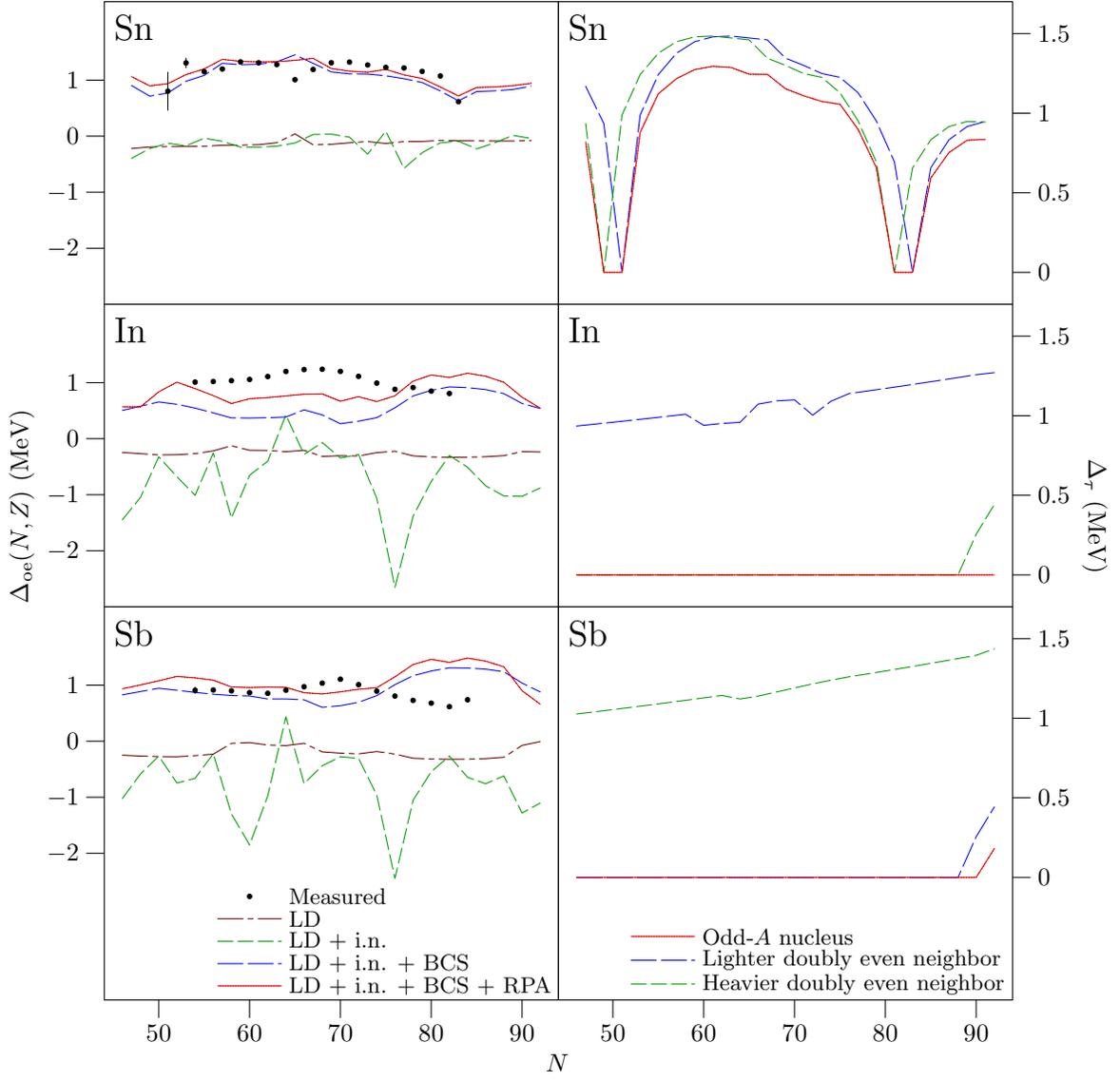}\par}
  \caption{\label{fig:oeSn}(Color online) Similar to
    Fig.~\ref{fig:oeN=Z} for the neighborhood of the Sn isotopes.}
\end{figure*}

Figure~\ref{fig:GSn} illustrates the need of both the nonzero $\alpha$
and the smaller $\bar G$. The quantities plotted in the upper left,
upper right, and lower right panels are the total calculated shell
correction $\delta E
= \delta E_\text{i.n.} + \delta E_\text{BCS} + \delta E_\text{RPA}$
and its empirical counterpart
$\delta E_\text{emp} = E_\text{emp} - E_\text{LD}$, where
$- E_\text{emp}$ is the measured binding energy. They are displayed
for the doubly even Sn isotopes as functions of~$N$. Different sets of
liquid drop parameters give rise to a difference of $\delta
E_\text{emp}$ between the panels. In the upper left panel, the pairing
parameters are inherited from the $N \approx Z$ region,
cf.~Eq.~\eqref{eq:GN=Z}. They describe fairly well the empirical
binding energies near the $N = 50$ shell closure but not at all near
the $N = 82$ shell closure. Because the Sn isotopes have constant
proton configuration, the $G_\tau$ that most significantly influences
the isotopic variation is $G_n$. When $\alpha$ is positive, $G_n$
decreases more with increasing $N$ than by the factor $A^{-0.7461}$.
The upper right panel shows the result when $\bar G = 7.196$~MeV is
kept---so that Eq.~\eqref{eq:GN=Z} would be retained for $N = Z$---but
$\alpha$ is set to 0.0170. Now $\delta E_\text{emp}$ is equally well
described at both shell closures, but the empirical
$\Delta_\text{oe}(N,50)$ is seen in the lower left panel to be vastly
overestimated. The top panel of of Fig.~\ref{fig:oeSn} shows that this
discrepancy is eliminated when $\bar G$ is reduced to $5.818$~MeV. As
seen from the lower right panel of Fig.~\ref{fig:GSn} this also
improves the reproduction of the measured doubly even binding energies
near both shell closures.

We notice in passing that, in particular, a discontinuity of the
measured two-neutron separation energy at $N = 66$ is reproduced.
Togashi\etal~\cite{ref:Tog18} describe this discontinuity as a second
order phase transition. In our calculations it is correlated with an
onset of oblate deformation at the entrance at $N = 68$ of the highly
degenerate $1h_{11/2}$ shell, cf. the appendix. This concurs with a
finding of Togashi\etal, based on an analysis of the result of a
large-scale shell model calculation, that these nuclei have oblate
deformations. In the upper panels of Fig.~\ref{fig:GSn}, the plots of
$\delta E$ behave differently at $N \approx 66$. Pairing thus
contributes to the formation of the discontinuity in our calculations.

Also shown in Fig.~\ref{fig:GSn} is the neutron-proton RPA energy
$E_\text{RPA,$np$}(N,50)$. It increases with increasing neutron excess
because the products in Eq.~\eqref{eq:UV} decrease with increasing
distance between $\lambda_n$ and $\lambda_p$. It is seen, however,
that in $^{142}$Sn with almost twice as many neutrons as protons, it
is only reduced numerically to about two thirds of its value in the
$N = Z$ nucleus $^{100}$Sn.

Figure~\ref{fig:oeSn} shows the measured and calculated odd-even mass
differences and the decompositions of the latter. The RPA contribution
to the calculated $\Delta_\text{oe}(N,Z)$ is positive with few
exceptions. On average it makes up 7, 31 and 14 per cent of the total
for the odd-$A$ isotopes of Sn, In and Sb. This dominantly positive
sign is qualitatively consistent with the values of
$\chi_{\tau\tau'}$. For $N=46$ they are approximately equal, about
3.3, and they decrease slightly to about 3.2 for $N = 54$. When $N$
increases further, $\chi_n$ increases to about 4.0 while $\chi_n$ and
$\chi_{np}$ continue decreasing to about 2.7 and 3.0, respectively.
That the $\chi_{\tau\tau'}$ of $^{100}$Sn are larger here than in the
calculation discussed in Sec.~\ref{sec:N=Z} is due to the smaller
$\bar G$.

Except for the largest $N$ we get $\Delta_\tau = 0$ when $N_\tau$ is
magic or magic $\pm$ 1. These are the cases when the Fermi level lies
within the magic gap in the single-nucleon spectrum. Otherwise
$\Delta_\tau > 0$. The emergence of $\Delta_p > 0$ in $^{90}$Sn,
$^{92}$Sn, and $^{92}$Sb reflects that $G_\text{cr,$p$}$ is close to
$G_p$ for the heaviest isotopes of In, Sn, and Sb. This is correlated
with low RPA contributions to the calculated $\Delta_\text{oe}(N,Z)$
in the isotopes of In and Sb with $N = 90$ and $92$.

\section{\label{sec:Zr}$^{102}\mathrm{Zr}$ region}

In the region around $^{102}$Zr we consider all doubly even and
odd-$A$ nuclei with $60 \le N \le 64$ and \linebreak
$38 \le Z \le 42$. As in the Sn region, we keep the $A$ exponent
$\zeta = -0.7461$ from Eq.~\eqref{eq:GN=Z} but adjust $\bar G$ and
$\alpha$ in Eq.~\eqref{eq:G} so as to reproduce the average of the
measured $\Delta_\text{oe}(N,Z)$ separately for odd $N$ and odd $Z$.
The result is
\begin{equation}\label{eq:GZr}
  G_\tau = 5.820 A^{-0.7461} (1 - 0.0132 M_T M_T') \text{ MeV} .
\end{equation}
Thus $\bar G$ is practically the same as in the Sn region,
cf.~Eq~\eqref{eq:GSn}, but $\alpha$ is significantly smaller.

\begin{figure*}
  {\centering \includegraphics{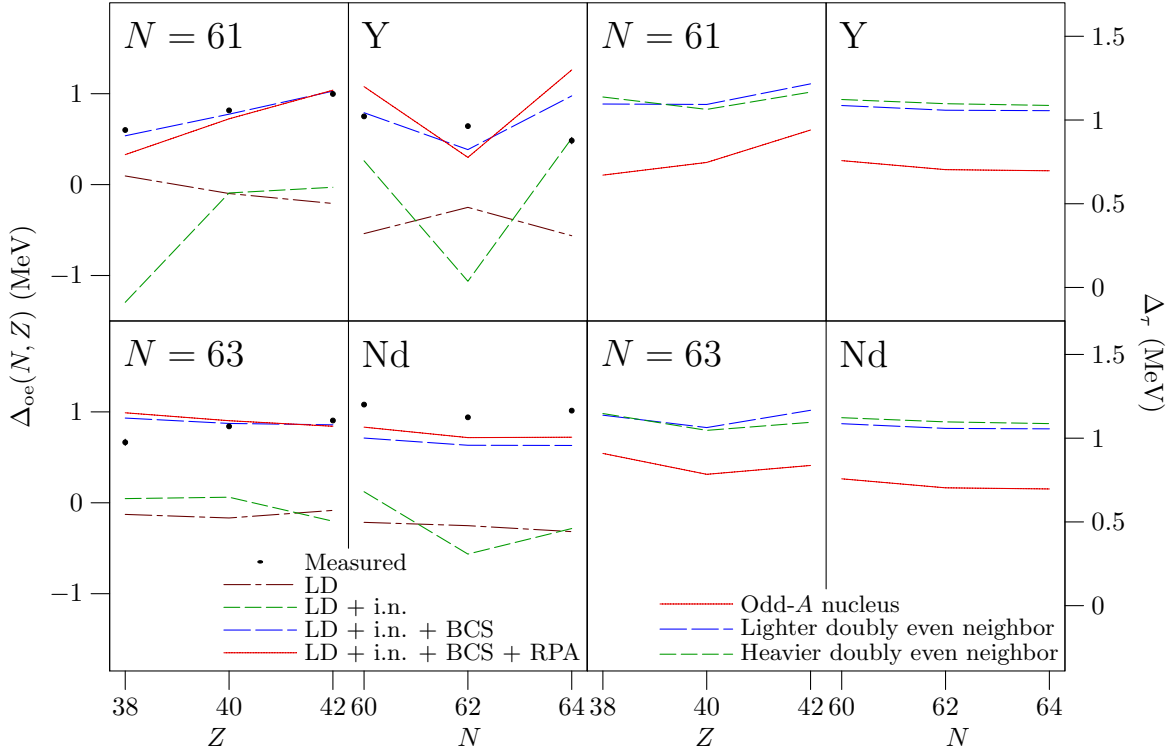}\par}
  \caption{\label{fig:oeZr}(Color online) Similar to
    Fig.~\ref{fig:oeN=Z} for the neighborhood of $^{102}$Zr.}
\end{figure*}
The measured and calculated odd-even mass differences are compared and
the decompositions of the latter shown in Fig.~\ref{fig:oeZr}. The
sign of the RPA contribution varies with a slight predominance of the
positive sign, which occurs in 8 out of 12 cases. This is consistent
with the values of $\chi_{\tau\tau'}$, which are $\chi_n \approx 3.4$
and $\chi_p \approx \chi_{np} \approx 3.2$. On average the RPA
contribution makes up 6\% of the total calculated
$\Delta_\text{oe}(N,Z)$.

The gap parameters $\Delta_\tau$ are almost constant with averages
about 1.1~MeV for even $N$ and $Z$ and 0.8~MeV for odd $A$. The latter
is close to the average of the calculated $\Delta_\text{oe}(N,Z)$.

\section{\label{sec:int}Interpolation}

We mentioned that the RPA energies $E_\text{RPA,$\tau\tau'$}$ are
interpolated across intervals of $G_{\tau\tau'}$ about the thresholds
$G_\text{cr,$\tau$}$ of BCS pairing to avoid the singularities there.
The interpolating function is the polynomial of third degree in
$G_{\tau\tau'}$ which joins the calculated values smoothly at the
interval endpoints. Interpolation is done for $\tau = \tau'$ and for
$\tau\tau' = np$ and $N = Z$. In terms of the interpolation width $w$
mentioned in Sec.~\ref{sec:LD}, the interval is
$G_\text{min,$\tau\tau'$} < G_{\tau\tau'} < G_\text{max,$\tau\tau'$}$
with
\begin{equation}\begin{gathered}
  G_\text{min,$\tau\tau'$}
    = (1 - w) \min (G_\text{cr,$\tau$},G_\text{cr,$\tau'$}) , \\
  G_\text{max,$\tau\tau'$}
    = (1 + w) \max (G_\text{cr,$\tau$},G_\text{cr,$\tau'$}) .
\end{gathered}\end{equation}
If $G_\text{max,$\tau\tau'$} = 0$ no
interpolation is done.

\begin{figure}
  {\centering\includegraphics{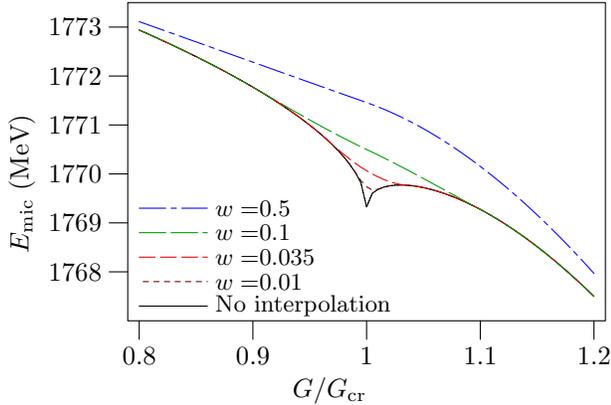}\par}
  \caption{\label{fig:ip}(Color online) Interpolated microscopic
    energy $E_\text{mic}$ as a function of $G/G_\text{cr}$ for
    different interpolation widths $w$ in the case of $^{100}$Sn. 
    See the text for details.}
\end{figure}
For even $N_\tau$ the threshold $G_\text{cr,$\tau$}$ increases with
increasing $\epsilon_{(N_\tau + 2) \tau} - \epsilon_{N_\tau \tau}$. It
is therefore particularly large when $N_\tau$ is magic. As a result
both $G_\text{cr,$\tau$}$ are close to the common value $G$ of $G_n$,
$G_p$, and $G_{np}$ in the doubly magic nuclei $^{56}$Ni and
$^{100}$Sn. For $^{100}$Sn, Fig.~\ref{fig:ip} shows the energy
$E_\text{mic}$ given by Eq.~\eqref{eq:mic} as a function of $G$ upon
interpolation with different $w$. A figure for $^{56}$Ni is very
similar. In this calculation we used the levels
$(\epsilon_{kn} +\epsilon_{kp})/2$ for both neutrons and protons so
that $G_\text{cr,$n$} = G_\text{cr,$p$} := G_\text{cr}$. It is seen
that the choice of $w$ can make a difference of 1--2~MeV in
$E_\text{mic}$ when $G_\text{cr}$ is close to $G$.

In Refs.~\cite{ref:Ben14,ref:Nee17}, $w = 0.5$ was chosen. This choice
was based on a comparison with a result of diagonalization of the
Hamiltonian~\eqref{eq:H} in a small valence space~\cite{ref:Ben13}.
Also Fig.~\ref{fig:bang} seems to suggest a fairly large interpolation
interval. In the latter calculation, however, the Hamiltonian is given
by Eq.~\eqref{eq:HBCS}, not Eq.~\eqref{eq:H}. Probably more
importantly, the single-nucleon levels are equidistant. The behavior
of the exact energy may be different when the Fermi level lies in a
gap in the single nucleon spectrum. In an early study, Feldman indeed
observed an approach of the exact result for the lowest excitation
energy to that of the RPA with increasing degeneracies of two separate
shells the lower of which is closed for $G = 0$~\cite{ref:Hog61}.
There is no way of determining the $w$ which best approximates the
exact minimum of any such Hamiltonian other than calibrating the
interpolation against an exact calculation, which is beyond our
capacity. Dukelsky\etal\ calculated the exact lowest energies for
isospin $T = 0$, 1 and 2 given by the Hamiltonian~\eqref{eq:H} in the
limit of isobaric invariance as functions of $G$ for the single
nucleus $^{64}$Ge with a different valence space and different
single-nucleon energies~\cite{ref:Duk06}, and even in this elaborate
calculation the dimension of the valence space (\textit{pf} shell plus
$1g_{9/2}$ subshell) is little greater than half of ours for
$^{56}$Ni.

With the large $w$ employed in Refs.~\cite{ref:Ben14,ref:Nee17}, quite
a few calculated binding energies depend on this parameter. This is
unsatisfactory because the choice of $w$ is largely arbitrary. We
prefer to trust the actual RPA energies unless there is a clear reason
not to do so. Such a reason is given by the observation that the exact
minimum of the Hamiltonian~\eqref{eq:H} must decrease as a function of
$G$ because the interaction is negative definite. As shown in
Fig.~\ref{fig:ip}, for the interpolated $E_\text{mic}$ of $^{100}$Sn
to similarly decrease as a function of $G$ it is necessary that
$w \gtrsim 0.035$. The same approximate limit results for $^{56}$Ni.
Therefore $w = 0.035$ was used in the present calculations.

\begin{figure}
  {\centering\includegraphics{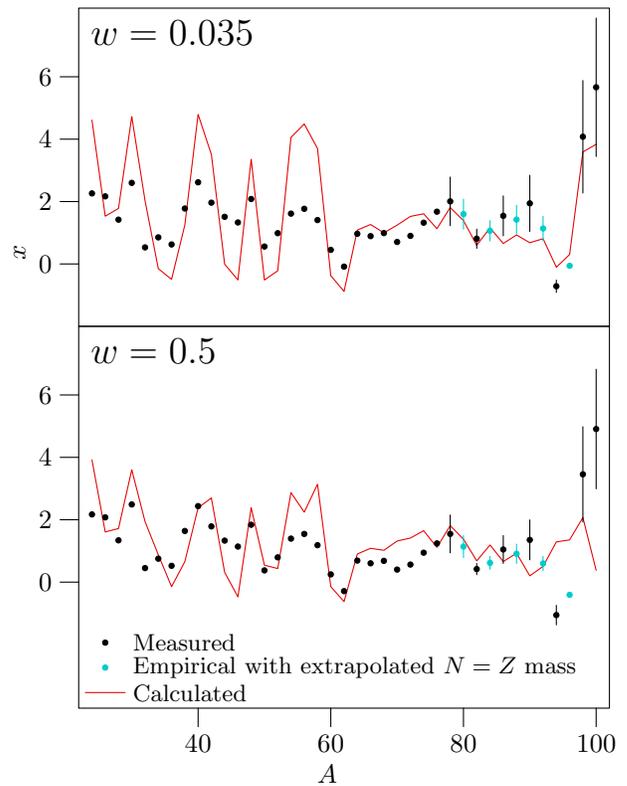}\par}
  \caption{\label{fig:x}(Color online) The calculated Wigner $x$
    as a function of $A$ for two different interpolation widths $w$ in
    comparison with the values extracted from mass data. In both
    calculations, the pair coupling constants $G_\tau$ are those of
    Sec.~\ref{sec:N=Z}. The liquid drop parameters are optimized
    separately for $w = 0.5$, resulting in
    $(a_v,a_{vt},a_s,a_{st},a_c) =
    (14.98,102.6,15.82,119.5,0.6400)$~MeV and rms deviations 0.735 and
    0.948~MeV from the doubly odd data and doubly even masses,
    respectively. The empirical points differ between the panels due
    to different $a_c$.}
\end{figure}
This diminishing of $w$ relative to the calculations in
Refs.~\cite{ref:Ben14,ref:Nee17} has implications for the
calculated\linebreak 'Wigner $x$', defined by~\cite{ref:Ben13}
\begin{equation}
  E (N,Z) = E_0 + \frac {|M_T|(|M_T| + x)} {2 \theta}
       + a_c \frac{Z(Z-1)}{A^{1/3}} B_c
\end{equation}
for a constant $A$ and $|M_T| = 0, 2, 4$ when $A \equiv 0 \mod 4$ and
$1, 3, 5$ when $A \equiv 2 \mod 4$. Here, besides $x$, also $E_0$ and
$\theta$ are constants. The value of $a_c$ is the one that results
from the fit of liquid drop parameters described in Sec.~\ref{sec:LD}.
As a function of $A$ the empirical $x$ has local maxima at the mass
numbers of the doubly magic nuclei $^{40}$Ca, $^{56}$Ni, and
$^{100}$Sn. This is seen in Fig.~\ref{fig:x} (and also in the plots of
$x$ in Refs.~\cite{ref:Ben14,ref:Nee17}, which resemble the 
bottom panel in Fig.~\ref{fig:x} in this respect) to be reproduced
with $w = 0.035$ but not with $w = 0.5$. The small $w$ is similarly
decisive for the sharpness of the calculated shell correction minimum
at $^{100}$Sn in the lower right panel of Fig.~\ref{fig:GSn}. These
successes of the small $w$ in reproducing qualitative features of the
patterns of binding energies near closed shells should evidently not
be seen as a proof that it best approximates the exact minimum of the
Hamiltonian~\eqref{eq:H}.

\section{\label{sec:sum}Summary}

The random-phase-approximation-amended (RPA-amended)
Nilsson-Strutinskij method of calculating nuclear binding energies was
reviewed in the form it has taken after modifications in the preceding
literature and in our present work. It was then applied in a study of
odd-mass nuclei. Three sets of such nuclei were considered. In terms
of the numbers $N$ and $Z$ of neutrons and protons and the mass number
$A = N + Z$ they are: (i) The sequence of nuclei with $Z = N - 1$ and
$25 \le A \le 99$; (ii) the odd-$A$ isotopes of In, Sn, and Sb with
$46 \le N \le 92$; (iii) the \linebreak odd-$A$ isotopes of Sr, Y, Zr,
Nb, and Mo with \linebreak $60 \le N \le 64$. An RPA based part of the
total shell correction which accounts for the pair-vibrational
correlation energy was found to contribute significantly to the
calculated odd-even mass differences, particularly in the light
nuclei. In the upper \textit{sd} shell it thus gives almost the entire
odd-even mass differences for odd $Z$ and about half of it for odd
$N$. In the heavier part of the set (i) it is less significant and the
contribution is negative for odd $N > 30$. In the sets (ii) and (iii)
it is dominantly positive and makes up 6--31\% of the total calculated
odd-even mass difference in various cases. These differences were
explained qualitatively in terms of a closed expression for a smooth
RPA counter term.

The coupling constants $G_n$, $G_p$, and $G_{np}$ of neutron, proton
and neutron-proton pairing interactions were expressed by
Eq.~\eqref{eq:G} in terms of parameters $\bar G$, $\zeta$, and
$\alpha$, which were set independently for regions of the chart of
nuclei each containing one of the sets (i)--(iii) of odd-$A$ nuclei.
In region (i), following previous studies of even-$A$ nuclei in this
region, we took $\alpha = 0$ and adjusted $\bar G$ and $\zeta$ to data
on doubly odd nuclei with $N = Z$. Remarkably, the resulting
parameters reproduce the typical size of the odd-even mass difference.
In the regions (ii) and (iii) the parameters $\bar G$ and $\alpha$
were fit directly to the odd-even mass differences with $\zeta$ kept
from region (i). Essentially the same $\bar G$ but different $\alpha$
resulted. The value of $\bar G$ derived from the data on doubly odd
$N = Z$ nuclei is 24\% greater than the one derived from odd-even mass
differences in the regions (ii) and (iii). As a result we got for
$^{100}$Sn, which belongs to both regions (i) and (ii), two values of
the common value of $G_n$, $G_p$, and $G_{np}$ differing by these
24\%. It was suggested that this difference be due to uncertainty of a
part of the data on doubly odd $N = Z$ nuclei.

An investigation of the binding energies of the Sn isotopes with even
$N$ showed that our model reproduces a discontinuity of the
two-neutron separation energy at $N = 66$ discussed recently by
Togashi\etal~\cite{ref:Tog18}. Like in their analysis of results of a
large-scale shell-model calculation, it is associated in our
calculation with an onset of oblate deformations at the entrance of
the $1h_{11/2}$ neutron shell. Pairing was found to contribute to the
formation of the discontinuity.

The RPA neutron-proton pair-vibrational correlation energy is expected
to decrease numerically with increasing neutron excess due to an
increasing mismatch of the occupations of single-neutron and
single-proton levels. In $^{142}$Sn, which has almost twice as many
neutrons as protons, it was found to be reduced anyway only to
about two thirds of its value in the $N = Z$ nucleus $^{100}$Sn.

The RPA-amended Nilsson-Strutinskij method involves an interpolation
of RPA energy terms across the thresholds of the pair coupling
constants for Bardeen-Cooper-Schrieffer pairing in the neutron or
proton system. Arguments were given for choosing the interpolation
interval substantially smaller than in previous applications of the
method, and such a smaller width was applied in our present
calculations. As a side effect, diminishing the width of the
interpolation interval resulted in an improved qualitative
correspondence between the variations with $A$ of the measured and
calculated `Wigner~$x$'.

\begin{acknowledgments}

  We would like to thank Stefan Frauendorf for providing access to the
  \textsc{tac} code that was used to calculate the deformations shown
  in the appendix and the corresponding single-nucleon levels used in
  this work.

\end{acknowledgments}

\appendix*

\section{Deformations}

Tables~\ref{tbl:defs} shows the deformations used in the calculations.
For odd $N = Z$ these are the deformations assumed for the lowest
states with $T \approx 0$.

{\squeezetable\begin{table*}
  \caption{\label{tbl:defs}Deformations used in the calculations.}
\begin{ruledtabular}
\begin{tabular}{
  D.{}{-1}D..{1.3}D..{2.0}D..{1.3}|D.{}{-1}D..{1.3}D..{2.0}D..{1.3}|
  D.{}{-1}D..{1.3}D..{2.0}D..{1.3}|D.{}{-1}D..{1.3}D..{2.0}D..{1.3}|
  D.{}{-1}D..{1.3}D..{2.0}D..{1.3}}
\multicolumn{1}{c}{Nucleus}&
\multicolumn{1}{c}{$\epsilon_2$}&
\multicolumn{1}{c}{$\gamma$ ($^\circ$)}&
\multicolumn{1}{c|}{$\epsilon_4$}&
\multicolumn{1}{c}{Nucleus}&
\multicolumn{1}{c}{$\epsilon_2$}&
\multicolumn{1}{c}{$\gamma$ ($^\circ$)}&
\multicolumn{1}{c|}{$\epsilon_4$}&
\multicolumn{1}{c}{Nucleus}&
\multicolumn{1}{c}{$\epsilon_2$}&
\multicolumn{1}{c}{$\gamma$ ($^\circ$)}&
\multicolumn{1}{c|}{$\epsilon_4$}&
\multicolumn{1}{c}{Nucleus}&
\multicolumn{1}{c}{$\epsilon_2$}&
\multicolumn{1}{c}{$\gamma$ ($^\circ$)}&
\multicolumn{1}{c|}{$\epsilon_4$}&
\multicolumn{1}{c}{Nucleus}&
\multicolumn{1}{c}{$\epsilon_2$}&
\multicolumn{1}{c}{$\gamma$ ($^\circ$)}&
\multicolumn{1}{c}{$\epsilon_4$}\\
\hline\\[-6pt]
^{24}.\text{O}&0.000&&0.000&^{50}.\text{Mn}&0.149&0&-0.005&
^{102}.\text{Sr}&0.249&60&-0.003&^{118}.\text{Cd}&0.116&60&0.009&
^{126}.\text{Sn}&0.000&&0.000\\
^{26}.\text{O}&0.000&&0.000&^{51}.\text{Mn}&0.111&0&-0.002&
^{78}.\text{Y}&0.222&60&0.013&^{120}.\text{Cd}&0.111&4&0.013&
^{127}.\text{Sn}&0.018&60&0.001\\
^{24}.\text{Ne}&0.091&0&0.000&^{52}.\text{Fe}&0.000&&0.000&
^{79}.\text{Y}&0.216&60&0.016&^{122}.\text{Cd}&0.084&29&0.012&
^{128}.\text{Sn}&0.000&&0.000\\
^{26}.\text{Ne}&0.000&&0.000&^{53}.\text{Fe}&0.077&0&0.004&
^{99}.\text{Y}&0.237&60&-0.015&^{124}.\text{Cd}&0.000&&0.000&
^{129}.\text{Sn}&0.016&60&0.001\\
^{28}.\text{Ne}&0.000&&0.000&^{54}.\text{Fe}&0.000&&0.000&
^{101}.\text{Y}&0.265&0&-0.003&^{126}.\text{Cd}&0.000&&0.000&
^{130}.\text{Sn}&0.000&&0.000\\
^{30}.\text{Ne}&0.000&&0.000&^{56}.\text{Fe}&0.000&&0.000&
^{103}.\text{Y}&0.241&60&0.000&^{128}.\text{Cd}&0.000&&0.000&
^{131}.\text{Sn}&0.019&0&0.006\\
^{24}.\text{Mg}&0.284&0&0.014&^{58}.\text{Fe}&0.000&&0.000&
^{80}.\text{Zr}&0.212&60&0.020&^{130}.\text{Cd}&0.000&&0.000&
^{132}.\text{Sn}&0.000&&0.000\\
^{25}.\text{Mg}&0.232&8&0.009&^{60}.\text{Fe}&0.000&&0.000&
^{81}.\text{Zr}&0.208&60&0.023&^{132}.\text{Cd}&0.000&&0.000&
^{133}.\text{Sn}&0.016&60&-0.004\\
^{26}.\text{Mg}&0.201&0&0.012&^{62}.\text{Fe}&0.043&60&0.001&
^{82}.\text{Zr}&0.204&60&0.025&^{134}.\text{Cd}&0.000&&0.000&
^{134}.\text{Sn}&0.000&&0.000\\
^{28}.\text{Mg}&0.000&&0.000&^{54}.\text{Co}&0.084&0&0.007&
^{84}.\text{Zr}&0.153&60&0.016&^{136}.\text{Cd}&0.000&&0.000&
^{135}.\text{Sn}&0.013&60&-0.003\\
^{30}.\text{Mg}&0.000&&0.000&^{55}.\text{Co}&0.049&0&0.004&
^{86}.\text{Zr}&0.000&&0.000&^{138}.\text{Cd}&0.000&&0.000&
^{136}.\text{Sn}&0.000&&0.000\\
^{32}.\text{Mg}&0.000&&0.000&^{56}.\text{Ni}&0.000&&0.000&
^{88}.\text{Zr}&0.000&&0.000&^{140}.\text{Cd}&0.093&0&-0.009&
^{137}.\text{Sn}&0.009&60&-0.001\\
^{34}.\text{Mg}&0.000&&0.000&^{57}.\text{Ni}&0.027&0&0.000&
^{90}.\text{Zr}&0.000&&0.000&^{95}.\text{In}&0.037&0&0.006&
^{138}.\text{Sn}&0.000&&0.000\\
^{26}.\text{Al}&0.223&30&0.002&^{58}.\text{Ni}&0.000&&0.000&
^{100}.\text{Zr}&0.249&0&-0.009&^{97}.\text{In}&0.028&0&0.006&
^{139}.\text{Sn}&0.000&&0.000\\
^{27}.\text{Al}&0.222&49&-0.005&^{60}.\text{Ni}&0.000&&0.000&
^{101}.\text{Zr}&0.259&0&-0.006&^{98}.\text{In}&0.040&0&0.009&
^{140}.\text{Sn}&0.000&&0.000\\
^{28}.\text{Si}&0.222&60&-0.003&^{62}.\text{Ni}&0.000&&0.000&
^{102}.\text{Zr}&0.265&0&-0.002&^{99}.\text{In}&0.022&0&0.005&
^{141}.\text{Sn}&0.012&0&0.000\\
^{29}.\text{Si}&0.122&60&0.002&^{64}.\text{Ni}&0.000&&0.000&
^{103}.\text{Zr}&0.266&0&0.005&^{101}.\text{In}&0.026&0&0.005&
^{142}.\text{Sn}&0.000&&0.000\\
^{30}.\text{Si}&0.000&&0.000&^{66}.\text{Ni}&0.000&&0.000&
^{104}.\text{Zr}&0.270&0&0.010&^{103}.\text{In}&0.036&0&0.005&
^{97}.\text{Sb}&0.033&60&-0.004\\
^{32}.\text{Si}&0.000&&0.000&^{58}.\text{Cu}&0.054&0&-0.001&
^{82}.\text{Nb}&0.203&60&0.025&^{105}.\text{In}&0.053&0&0.004&
^{99}.\text{Sb}&0.026&60&-0.004\\
^{34}.\text{Si}&0.000&&0.000&^{59}.\text{Cu}&0.039&0&0.000&
^{83}.\text{Nb}&0.200&60&0.027&^{107}.\text{In}&0.073&0&0.004&
^{101}.\text{Sb}&0.022&60&-0.004\\
^{36}.\text{Si}&0.000&&0.000&^{60}.\text{Zn}&0.000&&0.000&
^{101}.\text{Nb}&0.240&0&-0.006&^{109}.\text{In}&0.082&0&0.005&
^{103}.\text{Sb}&0.026&60&-0.004\\
^{38}.\text{Si}&0.132&0&-0.005&^{61}.\text{Zn}&0.010&0&0.000&
^{103}.\text{Nb}&0.257&0&0.000&^{111}.\text{In}&0.082&0&0.007&
^{105}.\text{Sb}&0.035&60&-0.005\\
^{30}.\text{P}&0.000&&0.000&^{62}.\text{Zn}&0.000&&0.000&
^{105}.\text{Nb}&0.263&10&0.011&^{113}.\text{In}&0.082&0&0.007&
^{107}.\text{Sb}&0.045&60&-0.005\\
^{31}.\text{P}&0.000&&0.000&^{64}.\text{Zn}&0.000&&0.000&
^{84}.\text{Mo}&0.200&60&0.031&^{115}.\text{In}&0.104&36&0.002&
^{109}.\text{Sb}&0.077&0&-0.012\\
^{32}.\text{S}&0.000&&0.000&^{66}.\text{Zn}&0.037&60&0.001&
^{85}.\text{Mo}&0.194&58&0.032&^{117}.\text{In}&0.113&41&0.005&
^{111}.\text{Sb}&0.083&0&-0.009\\
^{33}.\text{S}&0.033&60&0.001&^{68}.\text{Zn}&0.000&&0.000&
^{86}.\text{Mo}&0.080&60&0.003&^{119}.\text{In}&0.107&2&0.009&
^{113}.\text{Sb}&0.079&60&-0.007\\
^{34}.\text{S}&0.000&&0.000&^{70}.\text{Zn}&0.000&&0.000&
^{88}.\text{Mo}&0.000&&0.000&^{121}.\text{In}&0.099&7&0.011&
^{115}.\text{Sb}&0.110&60&-0.009\\
^{36}.\text{S}&0.000&&0.000&^{62}.\text{Ga}&0.011&0&0.000&
^{90}.\text{Mo}&0.000&&0.000&^{123}.\text{In}&0.080&7&0.011&
^{117}.\text{Sb}&0.126&60&-0.007\\
^{38}.\text{S}&0.000&&0.000&^{63}.\text{Ga}&0.002&0&0.000&
^{92}.\text{Mo}&0.000&&0.000&^{125}.\text{In}&0.051&0&0.008&
^{119}.\text{Sb}&0.128&60&-0.002\\
^{40}.\text{S}&0.000&&0.000&^{64}.\text{Ge}&0.000&&0.000&
^{94}.\text{Mo}&0.000&&0.000&^{127}.\text{In}&0.027&0&0.005&
^{121}.\text{Sb}&0.118&60&0.003\\
^{42}.\text{S}&0.000&&0.000&^{65}.\text{Ge}&0.101&0&0.004&
^{102}.\text{Mo}&0.219&26&0.001&^{129}.\text{In}&0.018&0&0.004&
^{123}.\text{Sb}&0.103&60&0.007\\
^{34}.\text{Cl}&0.054&60&0.003&^{66}.\text{Ge}&0.091&0&0.004&
^{103}.\text{Mo}&0.226&26&0.006&^{131}.\text{In}&0.014&0&0.004&
^{125}.\text{Sb}&0.085&60&0.008\\
^{35}.\text{Cl}&0.027&60&0.001&^{68}.\text{Ge}&0.113&60&0.002&
^{104}.\text{Mo}&0.241&21&0.005&^{133}.\text{In}&0.016&0&0.004&
^{127}.\text{Sb}&0.051&60&0.001\\
^{36}.\text{Ar}&0.000&&0.000&^{70}.\text{Ge}&0.121&60&0.005&
^{105}.\text{Mo}&0.251&17&0.007&^{135}.\text{In}&0.022&0&0.004&
^{129}.\text{Sb}&0.026&60&-0.002\\
^{37}.\text{Ar}&0.012&60&0.000&^{72}.\text{Ge}&0.000&&0.000&
^{106}.\text{Mo}&0.255&16&0.012&^{137}.\text{In}&0.033&0&0.003&
^{131}.\text{Sb}&0.017&60&-0.003\\
^{38}.\text{Ar}&0.000&&0.000&^{74}.\text{Ge}&0.000&&0.000&
^{86}.\text{Tc}&0.189&57&0.034&^{139}.\text{In}&0.056&0&0.000&
^{133}.\text{Sb}&0.014&60&-0.003\\
^{40}.\text{Ar}&0.000&&0.000&^{66}.\text{As}&0.114&0&0.007&
^{87}.\text{Tc}&0.022&60&-0.001&^{141}.\text{In}&0.086&0&-0.005&
^{135}.\text{Sb}&0.016&60&-0.003\\
^{42}.\text{Ar}&0.000&&0.000&^{67}.\text{As}&0.114&0&0.009&
^{88}.\text{Ru}&0.000&&0.000&^{96}.\text{Sn}&0.000&&0.000&
^{137}.\text{Sb}&0.022&60&-0.004\\
^{44}.\text{Ar}&0.000&&0.000&^{68}.\text{Se}&0.171&60&-0.002&
^{89}.\text{Ru}&0.005&0&0.000&^{97}.\text{Sn}&0.014&0&0.002&
^{139}.\text{Sb}&0.033&60&-0.005\\
^{46}.\text{Ar}&0.000&&0.000&^{69}.\text{Se}&0.153&60&0.000&
^{90}.\text{Ru}&0.000&&0.000&^{98}.\text{Sn}&0.000&&0.000&
^{141}.\text{Sb}&0.066&0&-0.016\\
^{38}.\text{K}&0.018&60&0.000&^{70}.\text{Se}&0.213&60&-0.002&
^{92}.\text{Ru}&0.000&&0.000&^{99}.\text{Sn}&0.022&0&0.005&
^{143}.\text{Sb}&0.096&0&-0.020\\
^{39}.\text{K}&0.008&60&0.000&^{72}.\text{Se}&0.200&60&0.002&
^{94}.\text{Ru}&0.000&&0.000&^{100}.\text{Sn}&0.000&&0.000&
^{98}.\text{Te}&0.000&&0.000\\
^{40}.\text{Ca}&0.000&&0.000&^{74}.\text{Se}&0.190&60&0.008&
^{96}.\text{Ru}&0.000&&0.000&^{101}.\text{Sn}&0.018&60&-0.002&
^{100}.\text{Te}&0.000&&0.000\\
^{41}.\text{Ca}&0.039&60&-0.005&^{76}.\text{Se}&0.000&&0.000&
^{98}.\text{Ru}&0.000&&0.000&^{102}.\text{Sn}&0.000&&0.000&
^{102}.\text{Te}&0.000&&0.000\\
^{42}.\text{Ca}&0.000&&0.000&^{78}.\text{Se}&0.058&0&0.000&
^{90}.\text{Rh}&0.008&0&0.000&^{103}.\text{Sn}&0.011&60&-0.001&
^{104}.\text{Te}&0.000&&0.000\\
^{44}.\text{Ca}&0.000&&0.000&^{70}.\text{Br}&0.244&60&-0.004&
^{91}.\text{Rh}&0.008&0&0.000&^{104}.\text{Sn}&0.000&&0.000&
^{106}.\text{Te}&0.000&&0.000\\
^{46}.\text{Ca}&0.000&&0.000&^{71}.\text{Br}&0.247&60&-0.003&
^{92}.\text{Pd}&0.000&&0.000&^{105}.\text{Sn}&0.000&&0.000&
^{108}.\text{Te}&0.000&&0.000\\
^{48}.\text{Ca}&0.000&&0.000&^{72}.\text{Kr}&0.273&60&-0.003&
^{93}.\text{Pd}&0.026&0&0.002&^{106}.\text{Sn}&0.000&&0.000&
^{110}.\text{Te}&0.000&&0.000\\
^{50}.\text{Ca}&0.000&&0.000&^{73}.\text{Kr}&0.247&60&-0.001&
^{94}.\text{Pd}&0.000&&0.000&^{107}.\text{Sn}&0.015&0&0.001&
^{112}.\text{Te}&0.000&&0.000\\
^{42}.\text{Sc}&0.066&60&-0.008&^{74}.\text{Kr}&0.248&60&0.001&
^{96}.\text{Pd}&0.000&&0.000&^{108}.\text{Sn}&0.000&&0.000&
^{114}.\text{Te}&0.000&&0.000\\
^{43}.\text{Sc}&0.047&60&-0.005&^{76}.\text{Kr}&0.220&60&0.008&
^{98}.\text{Pd}&0.000&&0.000&^{109}.\text{Sn}&0.009&60&-0.001&
^{116}.\text{Te}&0.111&60&-0.009\\
^{44}.\text{Ti}&0.000&&0.000&^{78}.\text{Kr}&0.201&60&0.014&
^{100}.\text{Pd}&0.000&&0.000&^{110}.\text{Sn}&0.000&&0.000&
^{118}.\text{Te}&0.132&60&-0.007\\
^{45}.\text{Ti}&0.023&60&-0.002&^{80}.\text{Kr}&0.063&0&0.001&
^{94}.\text{Ag}&0.032&0&0.004&^{111}.\text{Sn}&0.009&0&0.000&
^{120}.\text{Te}&0.132&60&-0.002\\
^{46}.\text{Ti}&0.000&&0.000&^{82}.\text{Kr}&0.051&0&0.002&
^{95}.\text{Ag}&0.019&0&0.003&^{112}.\text{Sn}&0.000&&0.000&
^{122}.\text{Te}&0.119&60&0.003\\
^{48}.\text{Ti}&0.000&&0.000&^{74}.\text{Rb}&0.231&60&0.002&
^{94}.\text{Cd}&0.000&&0.000&^{113}.\text{Sn}&0.029&0&0.001&
^{124}.\text{Te}&0.100&60&0.008\\
^{50}.\text{Ti}&0.000&&0.000&^{75}.\text{Rb}&0.229&60&0.004&
^{96}.\text{Cd}&0.000&&0.000&^{114}.\text{Sn}&0.000&&0.000&
^{126}.\text{Te}&0.076&60&0.008\\
^{52}.\text{Ti}&0.000&&0.000&^{76}.\text{Sr}&0.238&60&0.006&
^{97}.\text{Cd}&0.027&0&0.005&^{115}.\text{Sn}&0.068&60&-0.004&
^{128}.\text{Te}&0.000&&0.000\\
^{54}.\text{Ti}&0.000&&0.000&^{77}.\text{Sr}&0.227&60&0.009&
^{98}.\text{Cd}&0.000&&0.000&^{116}.\text{Sn}&0.000&&0.000&
^{130}.\text{Te}&0.000&&0.000\\
^{46}.\text{V}&0.046&0&-0.004&^{78}.\text{Sr}&0.218&60&0.013&
^{100}.\text{Cd}&0.000&&0.000&^{117}.\text{Sn}&0.061&60&-0.001&
^{132}.\text{Te}&0.000&&0.000\\
^{47}.\text{V}&0.114&0&-0.010&^{80}.\text{Sr}&0.205&60&0.018&
^{102}.\text{Cd}&0.000&&0.000&^{118}.\text{Sn}&0.092&60&0.000&
^{134}.\text{Te}&0.000&&0.000\\
^{48}.\text{Cr}&0.150&0&-0.014&^{82}.\text{Sr}&0.073&60&0.003&
^{104}.\text{Cd}&0.000&&0.000&^{119}.\text{Sn}&0.088&60&0.001&
^{136}.\text{Te}&0.000&&0.000\\
^{49}.\text{Cr}&0.148&0&-0.009&^{84}.\text{Sr}&0.000&&0.000&
^{106}.\text{Cd}&0.000&&0.000&^{120}.\text{Sn}&0.088&60&0.004&
^{138}.\text{Te}&0.000&&0.000\\
^{50}.\text{Cr}&0.100&0&-0.002&^{86}.\text{Sr}&0.000&&0.000&
^{108}.\text{Cd}&0.084&0&0.003&^{121}.\text{Sn}&0.083&60&0.006&
^{140}.\text{Te}&0.000&&0.000\\
^{52}.\text{Cr}&0.000&&0.000&^{98}.\text{Sr}&0.248&60&-0.019&
^{110}.\text{Cd}&0.086&0&0.005&^{122}.\text{Sn}&0.076&60&0.007&
^{142}.\text{Te}&0.000&&0.000\\
^{54}.\text{Cr}&0.000&&0.000&^{99}.\text{Sr}&0.266&0&-0.007&
^{112}.\text{Cd}&0.092&0&0.006&^{123}.\text{Sn}&0.064&60&0.007&
^{144}.\text{Te}&0.086&0&-0.014\\
^{56}.\text{Cr}&0.000&&0.000&^{100}.\text{Sr}&0.252&60&-0.013&
^{114}.\text{Cd}&0.122&60&-0.001&^{124}.\text{Sn}&0.039&60&0.003\\
^{58}.\text{Cr}&0.087&0&0.002&^{101}.\text{Sr}&0.251&60&-0.008&
^{116}.\text{Cd}&0.127&60&0.004&^{125}.\text{Sn}&0.019&60&0.001
\end{tabular}
\end{ruledtabular}
\end{table*}}

\bibliography{oe}

%merlin.mbs apsrev4-1.bst 2010-07-25 4.21a (PWD, AO, DPC) hacked
%Control: key (0)
%Control: author (8) initials jnrlst
%Control: editor formatted (1) identically to author
%Control: production of article title (-1) disabled
%Control: page (0) single
%Control: year (1) truncated
%Control: production of eprint (0) enabled
\begin{thebibliography}{33}%
\makeatletter
\providecommand \@ifxundefined [1]{%
 \@ifx{#1\undefined}
}%
\providecommand \@ifnum [1]{%
 \ifnum #1\expandafter \@firstoftwo
 \else \expandafter \@secondoftwo
 \fi
}%
\providecommand \@ifx [1]{%
 \ifx #1\expandafter \@firstoftwo
 \else \expandafter \@secondoftwo
 \fi
}%
\providecommand \natexlab [1]{#1}%
\providecommand \enquote  [1]{``#1''}%
\providecommand \bibnamefont  [1]{#1}%
\providecommand \bibfnamefont [1]{#1}%
\providecommand \citenamefont [1]{#1}%
\providecommand \href@noop [0]{\@secondoftwo}%
\providecommand \href [0]{\begingroup \@sanitize@url \@href}%
\providecommand \@href[1]{\@@startlink{#1}\@@href}%
\providecommand \@@href[1]{\endgroup#1\@@endlink}%
\providecommand \@sanitize@url [0]{\catcode `\\12\catcode `\$12\catcode
  `\&12\catcode `\#12\catcode `\^12\catcode `\_12\catcode `\%12\relax}%
\providecommand \@@startlink[1]{}%
\providecommand \@@endlink[0]{}%
\providecommand \url  [0]{\begingroup\@sanitize@url \@url }%
\providecommand \@url [1]{\endgroup\@href {#1}{\urlprefix }}%
\providecommand \urlprefix  [0]{URL }%
\providecommand \Eprint [0]{\href }%
\providecommand \doibase [0]{http://dx.doi.org/}%
\providecommand \selectlanguage [0]{\@gobble}%
\providecommand \bibinfo  [0]{\@secondoftwo}%
\providecommand \bibfield  [0]{\@secondoftwo}%
\providecommand \translation [1]{[#1]}%
\providecommand \BibitemOpen [0]{}%
\providecommand \bibitemStop [0]{}%
\providecommand \bibitemNoStop [0]{.\EOS\space}%
\providecommand \EOS [0]{\spacefactor3000\relax}%
\providecommand \BibitemShut  [1]{\csname bibitem#1\endcsname}%
\let\auto@bib@innerbib\@empty
%</preamble>
\bibitem [{\citenamefont {Bardeen}\ \emph
  {et~al.}(1957{\natexlab{a}})\citenamefont {Bardeen}, \citenamefont {Cooper},\
  and\ \citenamefont {Schrieffer}}]{ref:Bar57a}%
  \BibitemOpen
  \bibfield  {author} {\bibinfo {author} {\bibfnamefont {J.}~\bibnamefont
  {Bardeen}}, \bibinfo {author} {\bibfnamefont {L.~N.}\ \bibnamefont {Cooper}},
  \ and\ \bibinfo {author} {\bibfnamefont {J.~R.}\ \bibnamefont {Schrieffer}},\
  }\href@noop {} {\bibfield  {journal} {\bibinfo  {journal} {Phys. Rev.}\
  }\textbf {\bibinfo {volume} {106}},\ \bibinfo {pages} {162} (\bibinfo {year}
  {1957}{\natexlab{a}})}\BibitemShut {NoStop}%
\bibitem [{\citenamefont {Bardeen}\ \emph
  {et~al.}(1957{\natexlab{b}})\citenamefont {Bardeen}, \citenamefont {Cooper},\
  and\ \citenamefont {Schrieffer}}]{ref:Bar57b}%
  \BibitemOpen
  \bibfield  {author} {\bibinfo {author} {\bibfnamefont {J.}~\bibnamefont
  {Bardeen}}, \bibinfo {author} {\bibfnamefont {L.~N.}\ \bibnamefont {Cooper}},
  \ and\ \bibinfo {author} {\bibfnamefont {J.~R.}\ \bibnamefont {Schrieffer}},\
  }\href@noop {} {\bibfield  {journal} {\bibinfo  {journal} {Phys. Rev.}\
  }\textbf {\bibinfo {volume} {108}},\ \bibinfo {pages} {1175} (\bibinfo {year}
  {1957}{\natexlab{b}})}\BibitemShut {NoStop}%
\bibitem [{\citenamefont {Bohr}\ \emph {et~al.}(1958)\citenamefont {Bohr},
  \citenamefont {Mottelson},\ and\ \citenamefont {Pines}}]{ref:Boh58}%
  \BibitemOpen
  \bibfield  {author} {\bibinfo {author} {\bibfnamefont {A.}~\bibnamefont
  {Bohr}}, \bibinfo {author} {\bibfnamefont {B.~R.}\ \bibnamefont {Mottelson}},
  \ and\ \bibinfo {author} {\bibfnamefont {D.}~\bibnamefont {Pines}},\
  }\href@noop {} {\bibfield  {journal} {\bibinfo  {journal} {Phys. Rev.}\
  }\textbf {\bibinfo {volume} {110}},\ \bibinfo {pages} {936} (\bibinfo {year}
  {1958})}\BibitemShut {NoStop}%
\bibitem [{\citenamefont {Bogolyubov}(1958)}]{ref:Bog58}%
  \BibitemOpen
  \bibfield  {author} {\bibinfo {author} {\bibfnamefont {N.~N.}\ \bibnamefont
  {Bogolyubov}},\ }\href@noop {} {\bibfield  {journal} {\bibinfo  {journal}
  {Dokl. Akad. Nauk SSSR}\ }\textbf {\bibinfo {volume} {119}},\ \bibinfo
  {pages} {52} (\bibinfo {year} {1958})},\ \bibinfo {note} {[Sov. Phys. Dokl.
  \textbf{3}, 279 (1958)]}\BibitemShut {NoStop}%
\bibitem [{\citenamefont {Solov'yov}(1958{\natexlab{a}})}]{ref:Sol58a}%
  \BibitemOpen
  \bibfield  {author} {\bibinfo {author} {\bibfnamefont {V.~G.}\ \bibnamefont
  {Solov'yov}},\ }\href@noop {} {\bibfield  {journal} {\bibinfo  {journal}
  {Dokl. Akad. Nauk SSSR}\ }\textbf {\bibinfo {volume} {123}},\ \bibinfo
  {pages} {437} (\bibinfo {year} {1958}{\natexlab{a}})},\ \bibinfo {note}
  {[Sov. Phys. Dokl. \textbf{3}, 1176 (1958)]}\BibitemShut {NoStop}%
\bibitem [{\citenamefont {Solov'yov}(1958{\natexlab{b}})}]{ref:Sol58b}%
  \BibitemOpen
  \bibfield  {author} {\bibinfo {author} {\bibfnamefont {V.~G.}\ \bibnamefont
  {Solov'yov}},\ }\href@noop {} {\bibfield  {journal} {\bibinfo  {journal}
  {Dokl. Akad. Nauk SSSR}\ }\textbf {\bibinfo {volume} {123}},\ \bibinfo
  {pages} {652} (\bibinfo {year} {1958}{\natexlab{b}})},\ \bibinfo {note}
  {[Sov. Phys. Dokl. \textbf{3}, 1197 (1958)]}\BibitemShut {NoStop}%
\bibitem [{\citenamefont {Bang}\ and\ \citenamefont
  {Krumlinde}(1970)}]{ref:Ban70}%
  \BibitemOpen
  \bibfield  {author} {\bibinfo {author} {\bibfnamefont {J.}~\bibnamefont
  {Bang}}\ and\ \bibinfo {author} {\bibfnamefont {J.}~\bibnamefont
  {Krumlinde}},\ }\href@noop {} {\bibfield  {journal} {\bibinfo  {journal}
  {Nucl. Phys. A}\ }\textbf {\bibinfo {volume} {141}},\ \bibinfo {pages} {18}
  (\bibinfo {year} {1970})}\BibitemShut {NoStop}%
\bibitem [{\citenamefont {Richardson}(1963)}]{ref:Ric63}%
  \BibitemOpen
  \bibfield  {author} {\bibinfo {author} {\bibfnamefont {R.~W.}\ \bibnamefont
  {Richardson}},\ }\href@noop {} {\bibfield  {journal} {\bibinfo  {journal}
  {Phys. Lett.}\ }\textbf {\bibinfo {volume} {3}},\ \bibinfo {pages} {277}
  (\bibinfo {year} {1963})}\BibitemShut {NoStop}%
\bibitem [{\citenamefont {Bohm}\ and\ \citenamefont {Pines}(1953)}]{ref:Bom53}%
  \BibitemOpen
  \bibfield  {author} {\bibinfo {author} {\bibfnamefont {D.}~\bibnamefont
  {Bohm}}\ and\ \bibinfo {author} {\bibfnamefont {D.}~\bibnamefont {Pines}},\
  }\href@noop {} {\bibfield  {journal} {\bibinfo  {journal} {Phys. Rev.}\
  }\textbf {\bibinfo {volume} {92}},\ \bibinfo {pages} {609} (\bibinfo {year}
  {1953})}\BibitemShut {NoStop}%
\bibitem [{\citenamefont {Strutinskij}(1966)}]{ref:Str66}%
  \BibitemOpen
  \bibfield  {author} {\bibinfo {author} {\bibfnamefont {V.~M.}\ \bibnamefont
  {Strutinskij}},\ }\href@noop {} {\bibfield  {journal} {\bibinfo  {journal}
  {Yad. Fiz.}\ }\textbf {\bibinfo {volume} {3}},\ \bibinfo {pages} {614}
  (\bibinfo {year} {1966})},\ \bibinfo {note} {[Sov. J. Nucl. Phys. \textbf{3},
  449 (1966)]}\BibitemShut {NoStop}%
\bibitem [{\citenamefont {Neerg{\aa}rd}(2009)}]{ref:Nee09}%
  \BibitemOpen
  \bibfield  {author} {\bibinfo {author} {\bibfnamefont {K.}~\bibnamefont
  {Neerg{\aa}rd}},\ }\href@noop {} {\bibfield  {journal} {\bibinfo  {journal}
  {Phys. Rev. C}\ }\textbf {\bibinfo {volume} {80}},\ \bibinfo {pages} {044313}
  (\bibinfo {year} {2009})}\BibitemShut {NoStop}%
\bibitem [{\citenamefont {Bentley}\ \emph {et~al.}(2014)\citenamefont
  {Bentley}, \citenamefont {Neerg{\aa}rd},\ and\ \citenamefont
  {Frauendorf}}]{ref:Ben14}%
  \BibitemOpen
  \bibfield  {author} {\bibinfo {author} {\bibfnamefont {I.}~\bibnamefont
  {Bentley}}, \bibinfo {author} {\bibfnamefont {K.}~\bibnamefont
  {Neerg{\aa}rd}}, \ and\ \bibinfo {author} {\bibfnamefont {S.}~\bibnamefont
  {Frauendorf}},\ }\href@noop {} {\bibfield  {journal} {\bibinfo  {journal}
  {Phys. Rev. C}\ }\textbf {\bibinfo {volume} {89}},\ \bibinfo {pages} {034302}
  (\bibinfo {year} {2014})}\BibitemShut {NoStop}%
\bibitem [{\citenamefont {Neerg{\aa}rd}(2016)}]{ref:Nee16}%
  \BibitemOpen
  \bibfield  {author} {\bibinfo {author} {\bibfnamefont {K.}~\bibnamefont
  {Neerg{\aa}rd}},\ }\href@noop {} {\bibfield  {journal} {\bibinfo  {journal}
  {Phys. Rev. C}\ }\textbf {\bibinfo {volume} {94}},\ \bibinfo {pages} {054328}
  (\bibinfo {year} {2016})}\BibitemShut {NoStop}%
\bibitem [{\citenamefont {Neerg{\aa}rd}(2017)}]{ref:Nee17}%
  \BibitemOpen
  \bibfield  {author} {\bibinfo {author} {\bibfnamefont {K.}~\bibnamefont
  {Neerg{\aa}rd}},\ }\href@noop {} {\bibfield  {journal} {\bibinfo  {journal}
  {Nucl. Theor.}\ }\textbf {\bibinfo {volume} {36}},\ \bibinfo {pages} {195}
  (\bibinfo {year} {2017})}\BibitemShut {NoStop}%
\bibitem [{\citenamefont {Bentley}(2010)}]{ref:Ben10}%
  \BibitemOpen
  \bibfield  {author} {\bibinfo {author} {\bibfnamefont {I.}~\bibnamefont
  {Bentley}},\ }\href@noop {} {Ph.D. thesis},\ \bibinfo  {school} {University
  of Notre Dame} (\bibinfo {year} {2010})\BibitemShut {NoStop}%
\bibitem [{\citenamefont {Nilsson}\ \emph {et~al.}(1969)\citenamefont
  {Nilsson}, \citenamefont {Tsang}, \citenamefont {Sobiczewski}, \citenamefont
  {Szymanski}, \citenamefont {Wycech}, \citenamefont {Gustafson}, \citenamefont
  {Lamm}, \citenamefont {M\"oller},\ and\ \citenamefont {Nilsson}}]{ref:Nil69}%
  \BibitemOpen
  \bibfield  {author} {\bibinfo {author} {\bibfnamefont {S.~G.}\ \bibnamefont
  {Nilsson}}, \bibinfo {author} {\bibfnamefont {C.~F.}\ \bibnamefont {Tsang}},
  \bibinfo {author} {\bibfnamefont {A.}~\bibnamefont {Sobiczewski}}, \bibinfo
  {author} {\bibfnamefont {Z.}~\bibnamefont {Szymanski}}, \bibinfo {author}
  {\bibfnamefont {S.}~\bibnamefont {Wycech}}, \bibinfo {author} {\bibfnamefont
  {C.}~\bibnamefont {Gustafson}}, \bibinfo {author} {\bibfnamefont {I.~L.}\
  \bibnamefont {Lamm}}, \bibinfo {author} {\bibfnamefont {P.}~\bibnamefont
  {M\"oller}}, \ and\ \bibinfo {author} {\bibfnamefont {B.}~\bibnamefont
  {Nilsson}},\ }\href@noop {} {\bibfield  {journal} {\bibinfo  {journal} {Nucl.
  Phys. A}\ }\textbf {\bibinfo {volume} {131}},\ \bibinfo {pages} {1} (\bibinfo
  {year} {1969})}\BibitemShut {NoStop}%
\bibitem [{\citenamefont {Larsson}(1973)}]{ref:Lar73}%
  \BibitemOpen
  \bibfield  {author} {\bibinfo {author} {\bibfnamefont {S.~A.}\ \bibnamefont
  {Larsson}},\ }\href@noop {} {\bibfield  {journal} {\bibinfo  {journal} {Phys.
  Scr.}\ }\textbf {\bibinfo {volume} {8}},\ \bibinfo {pages} {17} (\bibinfo
  {year} {1973})}\BibitemShut {NoStop}%
\bibitem [{\citenamefont {Seeger}\ and\ \citenamefont
  {Howard}(1975)}]{ref:See75}%
  \BibitemOpen
  \bibfield  {author} {\bibinfo {author} {\bibfnamefont {P.~A.}\ \bibnamefont
  {Seeger}}\ and\ \bibinfo {author} {\bibfnamefont {W.~M.}\ \bibnamefont
  {Howard}},\ }\href@noop {} {\bibfield  {journal} {\bibinfo  {journal} {Nucl.
  Phys. A}\ }\textbf {\bibinfo {volume} {238}},\ \bibinfo {pages} {491}
  (\bibinfo {year} {1975})}\BibitemShut {NoStop}%
\bibitem [{\citenamefont {Edmonds}(1957)}]{ref:Edm57}%
  \BibitemOpen
  \bibfield  {author} {\bibinfo {author} {\bibfnamefont {A.~R.}\ \bibnamefont
  {Edmonds}},\ }\href@noop {} {\emph {\bibinfo {title} {Angular Momentum in
  Quantum Mechanics}}}\ (\bibinfo  {publisher} {Princeton University Press},\
  \bibinfo {address} {Princeton},\ \bibinfo {year} {1957})\BibitemShut
  {NoStop}%
\bibitem [{\citenamefont {Swiatecki}(1956)}]{ref:Swi56}%
  \BibitemOpen
  \bibfield  {author} {\bibinfo {author} {\bibfnamefont {W.~J.}\ \bibnamefont
  {Swiatecki}},\ }\href@noop {} {\bibfield  {journal} {\bibinfo  {journal}
  {Phys. Rev.}\ }\textbf {\bibinfo {volume} {104}},\ \bibinfo {pages} {993}
  (\bibinfo {year} {1956})}\BibitemShut {NoStop}%
\bibitem [{\citenamefont {Nilsson}(1955)}]{ref:Nil55}%
  \BibitemOpen
  \bibfield  {author} {\bibinfo {author} {\bibfnamefont {S.~G.}\ \bibnamefont
  {Nilsson}},\ }\href@noop {} {\bibfield  {journal} {\bibinfo  {journal} {Mat.
  Fys. Medd. Dan. Vid. Selsk.}\ }\textbf {\bibinfo {volume} {29}},\ \bibinfo
  {pages} {\#16} (\bibinfo {year} {1955})}\BibitemShut {NoStop}%
\bibitem [{\citenamefont {Bengtsson}\ and\ \citenamefont
  {Ragnarsson}(1985)}]{ref:Ben85}%
  \BibitemOpen
  \bibfield  {author} {\bibinfo {author} {\bibfnamefont {T.}~\bibnamefont
  {Bengtsson}}\ and\ \bibinfo {author} {\bibfnamefont {I.}~\bibnamefont
  {Ragnarsson}},\ }\href@noop {} {\bibfield  {journal} {\bibinfo  {journal}
  {Nucl. Phys. A}\ }\textbf {\bibinfo {volume} {436}},\ \bibinfo {pages} {14}
  (\bibinfo {year} {1985})}\BibitemShut {NoStop}%
\bibitem [{\citenamefont {Ring}\ and\ \citenamefont
  {Schuck}(1980)}]{ref:Rin80}%
  \BibitemOpen
  \bibfield  {author} {\bibinfo {author} {\bibfnamefont {P.}~\bibnamefont
  {Ring}}\ and\ \bibinfo {author} {\bibfnamefont {P.}~\bibnamefont {Schuck}},\
  }\href@noop {} {\emph {\bibinfo {title} {The Nuclear Many-Body Problem}}}\
  (\bibinfo  {publisher} {Springer, Berlin},\ \bibinfo {year}
  {1980})\BibitemShut {NoStop}%
\bibitem [{\citenamefont {Strutinsky}(1967)}]{ref:Str67}%
  \BibitemOpen
  \bibfield  {author} {\bibinfo {author} {\bibfnamefont {V.~M.}\ \bibnamefont
  {Strutinsky}},\ }\href@noop {} {\bibfield  {journal} {\bibinfo  {journal}
  {Nucl. Phys. A}\ }\textbf {\bibinfo {volume} {95}},\ \bibinfo {pages} {420}
  (\bibinfo {year} {1967})}\BibitemShut {NoStop}%
\bibitem [{\citenamefont {Hinohara}\ and\ \citenamefont
  {Nazarewicz}(2016)}]{ref:Hin16}%
  \BibitemOpen
  \bibfield  {author} {\bibinfo {author} {\bibfnamefont {N.}~\bibnamefont
  {Hinohara}}\ and\ \bibinfo {author} {\bibfnamefont {W.}~\bibnamefont
  {Nazarewicz}},\ }\href@noop {} {\bibfield  {journal} {\bibinfo  {journal}
  {Phys. Rev. Lett.}\ }\textbf {\bibinfo {volume} {116}},\ \bibinfo {pages}
  {152502} (\bibinfo {year} {2016})}\BibitemShut {NoStop}%
\bibitem [{\citenamefont {Vaquero}\ \emph {et~al.}(2013)\citenamefont
  {Vaquero}, \citenamefont {Egido},\ and\ \citenamefont
  {Rodr\'iguez}}]{ref:Vac13}%
  \BibitemOpen
  \bibfield  {author} {\bibinfo {author} {\bibfnamefont {N.~L.}\ \bibnamefont
  {Vaquero}}, \bibinfo {author} {\bibfnamefont {J.~L.}\ \bibnamefont {Egido}},
  \ and\ \bibinfo {author} {\bibfnamefont {T.~R.}\ \bibnamefont
  {Rodr\'iguez}},\ }\href@noop {} {\bibfield  {journal} {\bibinfo  {journal}
  {Phys. Rev. C}\ }\textbf {\bibinfo {volume} {88}},\ \bibinfo {pages} {064311}
  (\bibinfo {year} {2013})}\BibitemShut {NoStop}%
\bibitem [{\citenamefont {Audi}\ \emph {et~al.}(2012)\citenamefont {Audi},
  \citenamefont {Wang}, \citenamefont {Wapstra}, \citenamefont {Kondev},
  \citenamefont {MacCormick}, \citenamefont {Xu},\ and\ \citenamefont
  {Pfeiffer}}]{ref:Aud12}%
  \BibitemOpen
  \bibfield  {author} {\bibinfo {author} {\bibfnamefont {G.}~\bibnamefont
  {Audi}}, \bibinfo {author} {\bibfnamefont {M.}~\bibnamefont {Wang}}, \bibinfo
  {author} {\bibfnamefont {A.~H.}\ \bibnamefont {Wapstra}}, \bibinfo {author}
  {\bibfnamefont {F.~G.}\ \bibnamefont {Kondev}}, \bibinfo {author}
  {\bibfnamefont {M.}~\bibnamefont {MacCormick}}, \bibinfo {author}
  {\bibfnamefont {X.}~\bibnamefont {Xu}}, \ and\ \bibinfo {author}
  {\bibfnamefont {P.}~\bibnamefont {Pfeiffer}},\ }\href@noop {} {\bibfield
  {journal} {\bibinfo  {journal} {Ch. Phys. C}\ }\textbf {\bibinfo {volume}
  {36}},\ \bibinfo {pages} {1287, 1603} (\bibinfo {year} {2012})}\BibitemShut
  {NoStop}%
\bibitem [{\citenamefont {Huang}\ \emph {et~al.}(2017)\citenamefont {Huang},
  \citenamefont {Audi}, \citenamefont {Wang}, \citenamefont {Kondev},
  \citenamefont {Naimi},\ and\ \citenamefont {Xu}}]{ref:Hua17}%
  \BibitemOpen
  \bibfield  {author} {\bibinfo {author} {\bibfnamefont {V.~J.}\ \bibnamefont
  {Huang}}, \bibinfo {author} {\bibfnamefont {G.}~\bibnamefont {Audi}},
  \bibinfo {author} {\bibfnamefont {M.}~\bibnamefont {Wang}}, \bibinfo {author}
  {\bibfnamefont {F.~G.}\ \bibnamefont {Kondev}}, \bibinfo {author}
  {\bibfnamefont {S.}~\bibnamefont {Naimi}}, \ and\ \bibinfo {author}
  {\bibfnamefont {X.}~\bibnamefont {Xu}},\ }\href@noop {} {\bibfield  {journal}
  {\bibinfo  {journal} {Ch. Phys. C}\ }\textbf {\bibinfo {volume} {41}},\
  \bibinfo {pages} {030002, 030003} (\bibinfo {year} {2017})}\BibitemShut
  {NoStop}%
\bibitem [{ref()}]{ref:ENSDF}%
  \BibitemOpen
  \href@noop {} {}\bibinfo {note} {Evaluated Nuclear Structure Data File,
  https://www.nndc.bnl.gov/ensdf, data retrieved Nov. 1, 2018}\BibitemShut
  {NoStop}%
\bibitem [{\citenamefont {Togashi}\ \emph {et~al.}(2018)\citenamefont
  {Togashi}, \citenamefont {Tsunoda}, \citenamefont {Otsuka}, \citenamefont
  {Shimizu},\ and\ \citenamefont {Honma}}]{ref:Tog18}%
  \BibitemOpen
  \bibfield  {author} {\bibinfo {author} {\bibfnamefont {T.}~\bibnamefont
  {Togashi}}, \bibinfo {author} {\bibfnamefont {Y.}~\bibnamefont {Tsunoda}},
  \bibinfo {author} {\bibfnamefont {T.}~\bibnamefont {Otsuka}}, \bibinfo
  {author} {\bibfnamefont {N.}~\bibnamefont {Shimizu}}, \ and\ \bibinfo
  {author} {\bibfnamefont {M.}~\bibnamefont {Honma}},\ }\href@noop {}
  {\bibfield  {journal} {\bibinfo  {journal} {Phys. Rev. Lett.}\ }\textbf
  {\bibinfo {volume} {121}},\ \bibinfo {pages} {062501} (\bibinfo {year}
  {2018})}\BibitemShut {NoStop}%
\bibitem [{\citenamefont {Bentley}\ and\ \citenamefont
  {Frauendorf}(2013)}]{ref:Ben13}%
  \BibitemOpen
  \bibfield  {author} {\bibinfo {author} {\bibfnamefont {I.}~\bibnamefont
  {Bentley}}\ and\ \bibinfo {author} {\bibfnamefont {S.}~\bibnamefont
  {Frauendorf}},\ }\href@noop {} {\bibfield  {journal} {\bibinfo  {journal}
  {Phys. Rev. C}\ }\textbf {\bibinfo {volume} {88}},\ \bibinfo {pages} {014322}
  (\bibinfo {year} {2013})}\BibitemShut {NoStop}%
\bibitem [{\citenamefont {H\"ogaasen-Feldman}(1961)}]{ref:Hog61}%
  \BibitemOpen
  \bibfield  {author} {\bibinfo {author} {\bibfnamefont {J.}~\bibnamefont
  {H\"ogaasen-Feldman}},\ }\href@noop {} {\bibfield  {journal} {\bibinfo
  {journal} {Nucl. Phys.}\ }\textbf {\bibinfo {volume} {28}},\ \bibinfo {pages}
  {258} (\bibinfo {year} {1961})}\BibitemShut {NoStop}%
\bibitem [{\citenamefont {Dukelsky}\ \emph {et~al.}(2006)\citenamefont
  {Dukelsky}, \citenamefont {Gueorguiev}, \citenamefont {{Van Isacker}},
  \citenamefont {Dimitrova}, \citenamefont {Errea},\ and\ \citenamefont {{Lerma
  H.}}}]{ref:Duk06}%
  \BibitemOpen
  \bibfield  {author} {\bibinfo {author} {\bibfnamefont {J.}~\bibnamefont
  {Dukelsky}}, \bibinfo {author} {\bibfnamefont {V.~G.}\ \bibnamefont
  {Gueorguiev}}, \bibinfo {author} {\bibfnamefont {P.}~\bibnamefont {{Van
  Isacker}}}, \bibinfo {author} {\bibfnamefont {S.}~\bibnamefont {Dimitrova}},
  \bibinfo {author} {\bibfnamefont {B.}~\bibnamefont {Errea}}, \ and\ \bibinfo
  {author} {\bibfnamefont {S.}~\bibnamefont {{Lerma H.}}},\ }\href@noop {}
  {\bibfield  {journal} {\bibinfo  {journal} {Phys. Rev. Lett.}\ }\textbf
  {\bibinfo {volume} {96}},\ \bibinfo {pages} {072503} (\bibinfo {year}
  {2006})}\BibitemShut {NoStop}%
\end{thebibliography}%

\end{document}